\documentclass{cimento}

\usepackage{graphicx}
\usepackage{amsmath}
\usepackage{amssymb}
\usepackage{bbold}
\usepackage[english]{babel}

\title{Cosmic-Ray-modified and driven instabilities}
\author{A.~Marcowith\from{ins:x}\thanks{Alexandre.Marcowith@umontpellier.fr}}
\instlist{\inst{ins:x}Laboratoire Univers et Particules de Montpellier, Universit\'e de Montpellier/CNRS, place E. Bataillon, cc072, 34095 Montpellier, France}

\begin{document}

\maketitle

\begin{abstract}
These lectures address the effects of Cosmic Rays over macroinstabilities which develop in the interstellar medium and the microinstabilities the particles are able to trigger themselves. The lectures are centered on the derivation of linear growth rates but also discuss some numerical simulations addressing the issue of magnetic field saturation. A particular emphasis is made on the streaming instability, an instability driven by anisotropic cosmic-ray distributions.
\end{abstract}

\section{Introduction}
Cosmic Rays (CRs), i.e. charged energetic particles beyond thermal equilibrium are an important component of the interstellar medium (ISM) aside thermal gas (or plasma), magnetic fields, dust and radiation to which they roughly share a same amount of energy density $E_{\rm CR} \sim 1~\rm{eV/cm^3}$ \cite{ref:gab}. CRs have an energy spectrum scaling as $E^{-2.7}$ above a few GeV per nucleons (or a few GeV as CRs are mostly composed of protons). As the rest mass energy of a proton is $m_p c^2 \simeq 0.938$ GeV, the energy density is carried by mildly relativistic particles. If CRs contribute a lot to ISM energy density or pressure they are negligible in terms of particle density, typically $n_{\rm CR} \sim 10^{-10}~\rm{cm^{-3}}$. The above numbers are mean values in the ISM, however close or in CR sources (like supernova remnants) CR energy density can be enhanced by one to three orders of magnitude and can hence become largely dominant. This underlines the space and also time (because CR sources have a finite lifetime) intermittent injection of CRs in the ISM. This fact makes CRs sources peculiar places where CR are more inclined to provide feed back over the local ISM notably through a class of driven instabilities \cite{ref:mar21}. CRs mostly because of their large pressure are also able to provide some feed back over instabilities that exist independently as we will discuss below.

\section{Cosmic Rays as source of back-reaction and instabilities}\label{S:CRP}
In this section we provide some quantitative estimations of the energy content in the relativistic component of the ISM. Before proceeding we have to make one first step by starting a discussion about the different astrophysical sources of energy injection in the ISM.
\subsection{Dominant sources of energy in the ISM}
The ISM is an important place because this is where stars form. The question of the star formation rate in our Galaxy is still an open scientific issue. It is a very complex physical process because it is highly non-linear with multiple sources of energy feed back at almost all scales \cite{ref:mac}. \\
In this complex system CRs are extracted from the thermalised plasma by different processes still object of intense research. During these processes some electric fields are at play to port particles from energies $E_c \sim {3 \over 2} k_B T$ in a medium of temperature T to several times of this energy. These electric fields can be associated with waves, or are produced in charge separation regions at shocks (in the so called ramp). Electric fields can also be produced in special regions where non-ideal effects (finite conductivity) develop as is the case in magnetic reconnection. All in all, these complex systems are able to extract an usually small (but not always) fraction of their total energy and to inject it into a small number of particles allowing them to reach relativistic energies usually by the interplay of turbulence. \\
Let us examine the different sources of energy injection in the ISM more closely (see \cite{ref:mac} for a complete review). Driving mechanisms at the origin of turbulent motions in the ISM can be list as: galactic shear motions and instabilities linked to gravitation, massive star radiation and winds, HII regions expansion, supernova explosions, young stellar objects jets. In each of these processes a fraction of the mechanical energy at disposal can injected into CRs. Considering all these sources, the dominant source of turbulence injection is due to supernova (SN) explosion. One estimate is given by :
\begin{eqnarray}\label{Eq:edo}
    \dot{e}_{\rm SN} = {\sigma_{\rm SN} \eta_{\rm SN} E_{\rm SN} \over \pi R_d^2 H_d} &\simeq& 3~10^{-26}~\rm{erg \over cm^3~s} \times \\
  &&\left(\eta_{\rm SN} \over 0.1 \right)\left({E_{\rm SN} \over 10^{51}~\rm{erg/s}} \right)\left({\sigma_{\rm SN} \over 1\rm{SNu}}\right)  \left({H_d \over 100~\rm{pc}}\right)^{-1}\left({R_d \over 20~\rm{kpc}}\right)^{-2} \ . \nonumber 
\end{eqnarray}
Here $E_{\rm SN}$ is mechanical energy deposited during a SN explosion, $\eta_{\rm SN}$ is the efficiency of the energy transfer into ISM gas, $\sigma_{\rm SN}$ is the SN rate, with $1~\rm{SNu}= 1~\rm{SN} (100~\rm{yr^{-1}})$ $(10^{10} L_B/L_\odot)^{-1}$, where $L_B$ is the blue luminosity of the Galaxy in terms of solar luminosity. Finally $H_d$ and $R_d$ are the disc height and radius. \\

However one should retain that SN explosions are not the only sources of  turbulent energy in the ISM. Massive star winds can also contribute to a substantial fraction of the amount in Eq.\ref{Eq:edo} (see discussions in \cite{ref:pari}). Comparatively an HII regions expansion contribute to about $3\times 10^{-30} \rm{erg s^{-1} cm^{-3}}$, instabilities contribute to a few $10^{-29} \rm{erg s^{-1} cm^{-3}}$, young stellar objects winds about $2 \times 10^{-28} \rm{erg s^{-1} cm^{-3}}$. It is however important to keep in mind that even if the latter are subdominant in terms of energy injection they inject energy at different scales with respect to SN (of the order of 10-100 pc). To what concerns CRs the different processes can inject energetic particles at very different scales from thousands of AU to pc, even if these energetic particles {\it do not} contribute to the local CR spectrum they can still have some feed back over the dynamics of nearby ISM.

\subsection{Cosmic Ray energetics} 
\paragraph{Cosmic Ray pressure} If we consider from radioactive decay studies and/or secondary to primary ratio measurements that CRs stay about 15 millions years in our Galaxy at energies of a few GeV \cite{ref:yan} and consider that 10\% of the above $\dot{e}$ (Eq.\ref{Eq:edo}) is converted into CRs we end with a typical energy density into CRs of $E_{\rm CR} \sim 1~\rm{eV/cm^{-3}}$. The CR differential energy density scales as $E^{-1.7}$. This efficiency of 0.1 is confirmed by different studies of CR non-linear acceleration at shock fronts \cite{ref:cap}. It should be mentioned here that the residues of SN explosions like pulsars and/or X-ray binaries are probably among strong sources of CR leptons (electrons and/or positrons) if not contributors of hadrons as well. \\
Another way to derive the above result is to directly start from observations of the CR spectrum on Earth. Measurements give the CR flux $\phi(E)$. We have the CR flux, differential density, differential energy density, differential pressure successively (at energies $E < 1~\rm{PeV}$, we further neglect here the CR hardening around 200 GeV/N)
\begin{eqnarray}
\phi_{\rm CR, \odot}(E) &\simeq& 2~\rm{cm^{-2} s^{-1} st^{-1} \rm{GeV}^{-1}} \left({E \over 1~\rm{GeV}}\right)^{-2.7} \ , \\
N_{\rm CR, \odot}(E) &\simeq& {8 \pi \over \beta(E) c} ~\rm{cm^{-3}  \rm{GeV}^{-1}} \left({E \over 1~\rm{GeV}}\right)^{-2.7} \ , \\
E_{\rm CR, \odot}(E) &\simeq& {8 \pi\over \beta(E) c} ~\rm{{GeV \over cm^3~GeV}} \left({E \over 1~\rm{GeV}}\right)^{-1.7} \ , \\
P_{\rm CR, \odot}(E) &\simeq& {8 \pi  \over 3 \beta(E) c} ~~\rm{{GeV \over cm^3~GeV}} \left({E \over 1~\rm{GeV}}\right)^{-1.7} \ .
\end{eqnarray}
Where $\beta(E)= \sqrt{1-{1 \over (1+ E/mc^2)^2}}$. Integrating $E_{\rm CR, \odot}$ over the energy gives $E_{\rm CR, tot, \odot} \sim 1.2~{\rm eV/cm^3}$ (we have assumed $\beta =1$). \\

However, as pointed out above CRs have a strong time- and space-intermittent distribution. Inside or nearby sources the energy density imparted into CRs can be higher. Considering an energy CR distribution in $E^{-2}$ and 10\% of the kinetic shock energy imparted into CRs we have a total pressure $P_{\rm CR}$ of 
\begin{equation}\label{Eq:PCR}
    P_{\rm CR, tot}  \simeq 10^3~\rm{eV \over cm^3} \left({\xi_{\rm CR} \over 0.1} \right) \left({n_{\rm ISM} \over 1~\rm{cm^{-3}}}\right) \left({u_{\rm sh} \over 1000~\rm{km/s}}\right)^2 \ ,
\end{equation}
the shock speed is $u_{\rm sh}$. Fast shocks harbour a clear pressure excess in CRs with respect to the mean ISM. Notice that because the CR spectrum at shocks is harder the differential pressure per logarithm bandwidth $dP_{\rm CR}/d\log{E} \propto E^0$, i.e. the highest CR energies carry as much as pressure as low energy CRs. In fact in fast supernova remnant shocks the spectrum is often softer than $E^{-2}$ and the pressure is still (more slightly) dominated by the GeV band.\\

\paragraph{Cosmic Ray pressure gradients} An important quantity is the CR gradient along the main magnetic field lines. Let us construct a simplified model of CR escape from the Galaxy based on \cite{ref:bla}. CRs are released from sources in the disc following a distribution scaling as $E^{-2}$ with an injection rate $Q_{\rm CR}(E)$, the density of CR wandering in the Galaxy with a diffusion coefficient $\kappa(E)$ is 
\begin{equation}
    n_{\rm CR}(E, z=0) = {Q_{\rm CR}(E) H_h \over {2\pi R_d^2 \kappa(E)}} \ . 
\end{equation}
Here we consider that all CR sources are located in an infinitely thin disc and that CR freely escape the Galaxy at the height of the halo $z = H_h$. At a peculiar height the CR density is $n_{\rm CR}(z,E)= n_{\rm CR}(E, z=0)(1-{|z|\over H_h})$. We express the injection in terms of the CR luminosity in the disc
\begin{equation}\label{Eq:QCR}
    Q_{\rm CR}(E) = {L_{\rm CR} \over \pi R_d^2 \Lambda E_0^2} \left({E \over 1~\rm{GeV}}\right)^{-2} 
\end{equation}
where $\Lambda =\ln(1 \rm{PeV}/1 \rm{GeV})\sim 13.8$, $L_{\rm CR} = {E_{\rm CR} V_{\rm CR} \over t_{\rm res}} \simeq 10^{41}$ erg/s, $E_0=1$ GeV. The CR-occupied volume is larger than the disc volume, we have chosen $V_{\rm CR} = 10 V_d$. The differential CR pressure is $P_{\rm CR}(E) \simeq n_{\rm CR} {E \over 3}$. Its gradient is 
\begin{equation}
    \partial_z P_{\rm CR}(E) \simeq 4~10^{-24} \rm{eV \over cm^4 GeV} \left({E \over 1~\rm{GeV}}\right)^{-1-\alpha} \ .
\end{equation}
In order to obtain this result we have assumed $R_d=20$ kpc and a diffusion coefficient $\kappa \simeq 3~10^{28}~\rm{cm^2/s}~(E/1 \rm{GeV})^{\alpha}$.\\
In a CR precursor shock, considering CR gradients along a magnetic field oriented along the shock normal we find 
$\partial_z P_{\rm CR} = {P_{\rm CR} \over L_d}$, here the precursor size corresponds to the CR diffusive length $L_d = {\kappa \over u_{\rm sh}}$ for a diffusion coefficient $\kappa$. This length is obtained by balancing the diffusive $L^2/\kappa$ to the advection time $L/u_{\rm sh}$ towards the shock front. For fast shocks with particles diffusing at a Bohm rate (i.e. $\kappa = {1 \over 3} R_L c$ where $R_L= E/q B$ is the particle Larmor radius) we find 
\begin{equation}
    L_d \simeq 3.3~10^{16}~\rm{cm} \left({E \over 1~\rm{GeV}}\right) \left({B \over 1~\mu G}\right) \left({u_{\rm sh} \over 1000~\rm{km/s}}\right)^{-1} \ , (\rm{about~0.01~pc}) 
\end{equation}
So the CR pressure gradient as function of the energy is using Eq. \ref{Eq:PCR}
\begin{equation}
\partial_z P_{\rm CR}(E) \simeq 3~10^{-14}~\rm{{eV \over cm^4 GeV}}  \left({\xi_{\rm CR} \over 0.1} \right) \left({n_{\rm ISM} \over 1~\rm{cm^{-3}}}\right) \left({u_{\rm sh} \over 1000~\rm{km/s}}\right)^3 \left({E \over 1~\rm{GeV}}\right)^{-2} \left({B \over 1~\mu G}\right)^{-1} \ .
\end{equation}
It is clear that even if at shock precursor magnetic fields are enhanced by a factor 10-100 strong CR gradients can develop there.\\

\paragraph{Cosmic Ray currents} Another important quantity is the current carried by CRs. We provide here again some estimations. The differential escaping flux of CR from the galactic disc is $\phi_{\rm CR}(E) = -\kappa(E) \partial_z n_{\rm CR}(E)$ (along the galactic height z) and the associated differential current density is $J_{\rm CR}(E) = q \phi_{\rm CR}$ \cite{ref:bla}. Using the Eq. \ref{Eq:QCR}
\begin{equation}
    J_{\rm CR,esc}(E) \simeq 2.0 ~10^{-13}~\rm{{esu \over cm^{2}~s~GeV}} \left({E \over 1~\rm{GeV}}\right)^{-2} \ . 
\end{equation}
The integrated current density is $J_{\rm CR, esc, tot} \simeq 2.0 ~10^{-13}~\rm{{esu \over cm^{2}~s}}$.\\

At shocks, CRs carry a current $e n_{\rm CR} u_{\rm sh}$, the CR total density is derived from Eq.\ref{Eq:PCR} is about $n_{\rm CR} \sim 10^{-7}~\rm{cm}^{-3}$ giving
\begin{equation}
    J_{\rm CR, prec} \simeq 4.3~10^{-9}~\rm{{esu \over cm^{2}~s}} \left({\xi_{\rm CR} \over 0.1} \right) \left({n_{\rm ISM} \over 1~\rm{cm^{-3}}}\right) \left({u_{\rm sh} \over 1000~\rm{km/s}}\right)^2 \ .
\end{equation}
The differential current density is scaling as $E^{-2}$ like the density. 

\subsection{Cosmic-Ray feed back processes}
CR can trigger instabilities or back react over instabilities because 1) they can ionise matter and contribute to magnetic field-plasma coupling, 2) because of their pressure, they can modify the matter equation of state and the local sound speed 3) because of their CR pressure gradient which acts as a force in the momentum fluid equation (see Eq.\ref{Eq:CRMHD2} below), 4) because of their current, they can generate magnetic fields. These lectures pay a particular attention to feed back processes 3 and 4 in Sect. \ref{S:MICR} but process 2 is briefly discussed in Sect. \ref{S:INST}. Discussion of process 1 is beyond the scope of the present lectures (see \cite{ref:gab} for more details).

\section{Model Equations in instability studies and general statements about plasma instabilities}
\subsection{The Boltzmann - Maxwell system}
In order to investigate instabilities in fluid dynamics one has at our disposal several mathematical frameworks aiming at an accurate description of the plasma state. It would take too much room to derive these system equations from the most fundamental ones, namely the Liouville or the Klimontovich equations. The reader can refer to the monograph by \cite{ref:nic} \S 3 \& 4 for more details.  \\

One of the most fundamental equation in dynamical systems is the Boltzmann equation or in case collisions can be neglected the Vlasov equation. It describes the phase space ($\vec{r},\vec{p}$) evolution of the particle distribution function $F(\vec{r},\vec{p},t)$, namely at a time t the number of particles in the phase space element $d^3\vec{r} d^3\vec{p}$ is $F(\vec{r},\vec{p},t)d^3\vec{r} d^3\vec{p}$. A plasma is composed of different species $a$, then a Boltzmann equation per species is necessary to describe their evolution. As charged particles, species $a$ also modify the electromagnetic fields hence the full system of model equation involves Boltzmann and Maxwell equations:

\begin{eqnarray}
 \partial_t F_a (\vec{r},\vec{p},t) + \vec{v}. \vec{\nabla} F_a + q_a \left(\vec{E} + {\vec{v} \over c} \wedge \vec{B}\right).\partial_{\vec{p}} F_a = \partial_t F_a(\vec{r},\vec{p},t)|_c && [Boltzmann] \label{Eq:Vlasov}\\
\vec{\nabla}.\vec{E} = 4\pi \sum_a q_a \int d^3\vec{p} ~F_a(\vec{r},\vec{p},t) + 4\pi \rho_{\rm c,ext} && [Gauss]  \label{Eq:Gauss}\\
\vec{\nabla} \wedge \vec{B} = {1 \over c} \partial_t \vec{E}+ {4\pi \over c} \sum_a q_a \int d^3\vec{p}~\vec{v} F_a + {4\pi \over c} \vec{J}_{\rm ext} && [Amp\grave{e}re] \label{Eq:Ampere}\\
\vec{\nabla} \wedge \vec{E} = -{1 \over c} \partial_t \vec{B} && [Faraday] \label{Eq:Faraday}\\
\vec{\nabla}.\vec{B} = 0 && [Thomson] \label{Eq:Thomson}
\end{eqnarray}
The RHS term of the Boltzmann equation describes the time effect of collisions. It reduces to the Vlasov equation if $\partial_t F_a(\vec{r},\vec{p},t)|_c =0$. We have also added the possibility to have external sources of charges $\rho_{\rm c, ext}$ and currents $\vec{J}_{\rm ext}$.

\subsection{Magneto-hydrodynamics}
From the Boltzmann equation it is possible by taking the moment over $\vec{p}$ to derive the fluid equations describing the evolution of an ensemble of particles in an electromagnetic field. These combined with Maxwell equations lead to the equation of magnetohydrodynamics (MHD) under standard assumptions: 1) characteristic time much larger than ion gyroperiod and mean free path time,
2) characteristic scale much larger than ion gyroradius and mean free path length. Namely, for non-relativistic flows they involve mass density conservation $\rho$ and an equation describing the momentum evolution $\rho \vec{u}$ 
\begin{eqnarray}
\partial_t \rho + \vec{\nabla}.(\rho \vec{u}) &=&  0 \ , \label{Eq:RHO} \\
\rho \left(\partial + \vec{u}.\vec{\nabla} \right) \vec{u} &=& {1 \over c} \vec{J} \wedge \vec{B} - \vec{\nabla} P \ , \label{Eq:EULER}\\
\vec{J} &=& {c \over 4\pi} \vec{\nabla} \wedge \vec{B}  \ ,  \\
{1 \over c} \partial_t \vec{B} &=& -\vec{\nabla} \wedge \vec{E} \ ,  \\
\vec{\nabla}.\vec{B} &=& 0  \ , \\
\vec{E} &=& -{\vec{u} \over c} \wedge \vec{B} \ . \label{Eq:OHM}
\end{eqnarray}
The last equation is the ideal Ohm's law. P is the gas (or plasma) pressure. These equations have to be complemented with an equation for the energy density of the gas or a relation linking the gas pressure and mass density through an equation of state, e.g. in case of a perfect gas described by its adiabatic index $\gamma_g$ we have $P=K \rho^{\gamma_g}$, K is a constant.

\subsection{Cosmic-Ray-magneto-hydrodynamics}
As for the question addressed in these lectures of the contribution of CRs to the production of electromagnetic fluctuations in a plasma, MHD equations have to be upgraded to account for the effects of CRs. These can modify flow dynamics because of their pressure $P_{\rm CR}$ and because of their current $\vec{J}_{\rm CR}$ which can both intervene in the momentum equation, Eq.\ref{Eq:EULER}. In case an energy equation is added for the gas, it is possible to add an energy equation for the "CR gas" by integrating the Vlasov over $\int d^3\vec{p} E(p)$, where $E(p)$ is the particle total energy. This results in the so-called bi-fluid description including a CR gas. The drawback of this formalism is that the information over the particle distribution in momentum is lost, so no kinetic effect associated with CRs can be retained. However, this method allows investigation of the back-reaction of a CR population over the gas dynamics and has been proved to be really useful in instability studies as described in Sect.\ref{S:INST}. We report here for completeness the governing bi-fluid equations \cite{ref:dcm, ref:tho, ref:but}:
\begin{eqnarray}
\partial_t \rho + \vec{\nabla} (\rho \vec{u}) &=&  0\ , \label{Eq:CRMHD1} \\
\partial_t (\rho \vec{u}) + \vec{\nabla}.\left (\rho \vec{u}\vec{u}+ P_{\rm tot} \bar{\bar{I}}-{\vec{B} \vec{B} \over 4\pi} \right) &=& \rho \vec{g}   \label{Eq:CRMHD2} \\
\partial_t e_g + \vec{\nabla}. (e_g \vec{u}) &=& -P_g \vec{\nabla}.\vec{u} -L + H + H_{\rm CR} \label{Eq:CRMHD3} \\
\partial_t e_{\rm CR} + \vec{\nabla}. \vec{F}_{\rm CR} &=& -P_{\rm CR} \vec{\nabla}.\vec{u} - H_{\rm CR}\ , \label{Eq:CRMHD4}
\end{eqnarray}
$e_g$ and $e_{\rm CR}$ are the energy density in the gas and in the CRs respectively. The total pressure is the sum of gas, magnetic field and CR contributions: $P_{\rm tot}= P_g + P_m + P_{\rm CR}$. The terms $L, H, H_{\rm CR}$ are the cooling and heating terms for the gas and the heating term due to CR streaming respectively. $\vec{F}_{\rm CR}$ is the CR flux it reads
\begin{equation}\label{Eq:FCR}
   \vec{F}_{\rm CR} = \vec{u} e_{\rm CR} + \vec{u}_{s} \left(e_{\rm CR} + P_{\rm CR}\right) - \kappa \vec{b} (\vec{b}.\vec{\nabla} e_{\rm CR}) \ .
\end{equation}
 The CR flux results from a contribution of three terms: CR advection with the gas (first RHS term in Eq. \ref{Eq:FCR}), CR streaming with respect to the gas (second RHS term) and spatial diffusion due to interaction magnetic irregularities (third RHS term). $\kappa$ is the CR diffusion coefficient parallel to the mean magnetic field and $\vec{b}= {\vec{B} \over B}$. We will describe the CR streaming process in Sect.\ref{S:LINEM}. The streaming speed is $\vec{u}_s = -{\vec{b}.\vec{\nabla}e_{\rm CR} \over |\vec{b}.\vec{\nabla}e_{\rm CR}|} $ and $H_{\rm CR} = |\vec{u}_s.\vec{\nabla} P_{\rm CR}|$. The CR fluid is described by its own adiabatic index $\gamma_{\rm CR}$.

\subsection{Instabilities in plasmas: general statements}
Quite generally speaking an instability results from an exponential growth of a wave in some peculiar modes (or wave number k) of a plasma. Before considering CRs and their relation to plasma instabilities it is important to discuss even succinctly what do we mean by a plasma instability. Plasma instabilities is a rich part of plasma physics. It involves many complex processes. Some classification of plasma instabilities can be found in \cite{ref:mik, ref:cap0}. One important aspect of the classification is the distinction among macroinstabilities and microinstabilities. 

\subsubsection{Macroinstabilities} Macroinstabilities involved a change  of the medium in the configuration space. Well known macroinstabilities are the Kelvin-Helmoltz or the Rayleigh-Taylor instability part of the magnetohydrodynamic (MHD) instabilities. In these lectures we will discuss one of these instabilities: the Parker-Jeans instability (see Sect. \ref{S:INST}) and how it is modified by the presence of CRs. 

\subsubsection{Microinstabilities} Microinstabilities usually involved a change of the medium in the velocity space. Almost all instabilities addressed here are microinstabilities because almost all are connected to anisotropy in the velocity distribution of CRs (see Sect. \ref{S:MICR}). Sect. \ref{S:LINEM} addresses in some details a famous microinstability, i.e. the streaming instability. 

\subsubsection{Instability criterion}
Once the equilibrium function of the particles in plasma is known it is possible to anticipate if it can lead or not to an instability. The method to evaluate the stability of a plasma system is the Nyquist method. It leads to a criterion for an instability which is the Penrose criterion. It is not the place here to discuss these two aspects of plasma stability studies in details. The interested reader is invited to read \cite{ref:kra} \S 9. We propose below a short formulation. 

\subsubsection{How to derive a growth rate ?}
The process has several steps.
\begin{enumerate}

\item Select a model equation. This means that you consider a model to describe a natural phenomenon. Hence depending on your choice you will have access to a given level of information and generality.

\item Properly define the unperturbed system: geometry, boundary conditions ... Often a difficult task. It can also have an impact over your result.  

\item Write your equations in terms of perturbed quantities. So each variables $A$ entering in your equations is developed as $A= A_0 + \delta A$, with $A_0$ the unperturbed part and $\delta A \ll A_0$ the perturbed one. The linear analysis allows to drop non-linear terms (because of higher orders in $\delta A$). 

\item Write your perturbed quantities in a Fourier form $\delta A \propto \exp(i (kx-\omega t))$. 

\item It often appears that you end with a system of coupled equations, then the need for a matrix analysis. The matrix eigenvalue analysis leads to an equation linking $\omega$ with $k$: this is the linear dispersion relation.

\item Solve for the imaginary part of $\omega(k)= {\cal R}(\omega)+i {\cal I}(\omega)$: ${\cal I}(\omega) \equiv \omega_I > 0$ leads to an instability, $\omega_I < 0$ leads to a decaying mode. The condition $\omega_I > 0$ is the instability criterion, it fixes a subspace in k where the system is unstable.

\end{enumerate}

\section{Cosmic-Ray-modified magneto-hydrodynamic instabilities}\label{S:INST}

We first present a broad overview of plasma instabilities in connection with the presence of Cosmic Rays including some rapid review of the literature. As stated in Sect.\ref{S:CRP} CRs as non-thermal particles are usually subdominant in density with respect to the background (thermal) plasma, i.e. $n_{\rm CR}/n_p \ll 1$, but they can carry as much as energy density (or pressure), i.e. $E_{\rm CR} = n_{\rm CR} m_p \langle (\gamma -1 )\rangle c^2$ (we assume otherwise specified that CRs are mostly relativistic protons). Because they carry a lot of momentum CRs are able to transfer through the generation of waves a fraction of this momentum to the background gas. The triggering of waves is at the heart of the driving of microinstabilities (see Sect.\ref{S:MICR}). 

\subsection{Main magneto-hydrodynamic instabilities}
Since CRs carry as much as pressure as the background gas and/or magnetic field they can modify the growth rate of the main fluid instabilities.  
\begin{itemize}
\item \underline{The Kelvin-Helmoltz instability}. The Kelvin-Helmoltz instability (KHI) is produced by a velocity shear in a flow \cite{ref:cha}. In astrophysics this situation occurs for instance in jets and outflows produced in young stellar objects, compact objects, in galactic winds or at the interface of interstellar and intergalactic media. CR effects over this instability are treated in \cite{ref:suz}.

\item \underline{The Rayleigh-Taylor and Parker instabilities}. The Rayleigh-Taylor instability (RTI) occurs when a plasma has a density contrast, with an heavier plasma above (in the inverse sense of a gravitation force) a lighter one, so with a density gradient in the same direction as the gravitational force \cite{ref:cha}. The latter can be replaced by an acceleration effect in moving fluids like the instability developing at the contact discontinuity separating forward (FS) and reverse shocks (RS) in supernova remnants (SNRs). CR effects are not explicitly treated in a specific work but usually treated in the context of supernova remnant as in \cite{ref:ryu92}. The Parker instability (PKI) is of a close nature and is described below.

\item \underline{The Thermal instability}. The so-called interstellar medium (ISM) are regions of the ISM stable with respect to the thermal instability  (THI) which is at the origin of the different ISM phases (stable thermodynamical conditions in the ISM). CR effects are treated in \cite{ref:sha}.

\item \underline{The Magneto-Rotational instability}. The magnetorotational instability (MRI) is an instability which develops in magnetised gas in differential rotation \cite{ref:cha60, ref:bal, ref:bal03}. It allows to transfer angular momentum outwardly and permits accretion. CR effects are treated for instance in \cite{ref:kuw15}.

\end{itemize}
Due to the size limitation of these lectures we have selected one particular instability among those and exemplify the effect of CR pressure. We have selected the Parker-Jeans instability which is one of the most important instabilities likely connected to the dynamo of the galactic magnetic field. The results concerning the other instabilities will be presented elsewhere. 

\subsection{The Parker-Jeans instability}
The Parker instability \cite{ref:par} occurs in a uniform disk of gas which is coupled to a magnetic field that is parallel to the disk while gravity applies in the vertical direction. The disk is in dynamical equilibrium under the balance of gravity and pressure (thermal and magnetic). Now consider a small perturbation which causes the magnetic field lines to oscillate around the equilibrium solution. Because of gravity, the gas loaded onto the field lines tends to slide off the peaks and to sink into the valleys. 
The increase of mass loads in the valleys makes them sink further, while the magnetic pressure causes the peaks to rise (buoyancy) as their mass load decreases. Hence, the initial perturbation is amplified. Parker \cite{ref:par} includes the contribution of CR pressure (without any diffusion). He finds that a criterion for the instability to grow is (L is the system typical size)
\begin{equation}\label{Eq:KCRIT}
    k_{\rm crit, P}^2 L^2 < {(P_g+P_m + P_{\rm CR})(P_g+P_m + P_{\rm CR}-\gamma_g) \over 2 P_{\rm CR} P_g \gamma_g} - {1 \over 4} \ .
\end{equation}
The instability hence develops at small wave numbers (large scales). 
\subsection{Cosmic Ray feed back over the Parker instability}
\begin{figure}
    \centering
    \includegraphics[width=\linewidth]{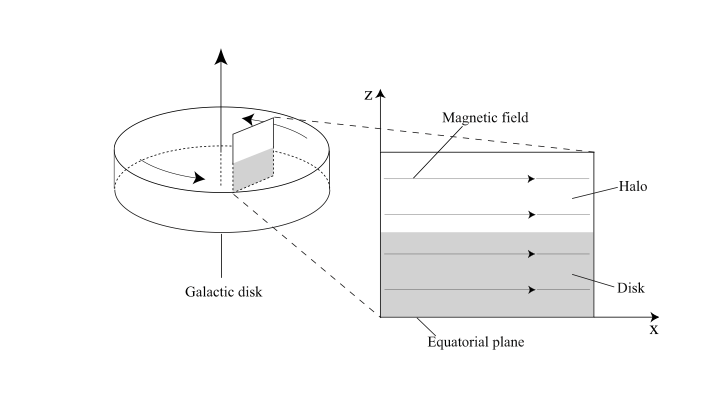}
    \caption{Sketch of the model. By courtesy of \cite{ref:ku04}.}
    \label{F:SPar}
\end{figure}

After liminar Parker's publication another linear analysis of the Parker instability including CR is due to \cite{ref:kuz}. Further analysis have been proposed by \cite{ref:ku04, ref:ku06, ref:ku20}. Ryu et al \cite{ref:ryu} conduct a similar study without including rotation and self-gravitation but including perpendicular CR diffusion coefficient. They show that the perpendicular diffusion coefficient has no strong impact over the instability growth. Kuwabara et al \cite{ref:ku04} only include parallel diffusion coefficient, rotation but no self-gravitation. The latter work is interesting to understand the effect of CRs over the Parker instability only (no self-gravitation so no Jeans instability). In order to conduct their analysis the authors considered a uniformly rotating galactic disk with an angular velocity $\vec{\Omega}$. The analysis is performed in 2D extracting Cartesian coordinates from a cylindrical coordinate attached to the disc ($-\vec{e}_r, \vec{e}_\phi, \vec{e}_z)$. Hence the Cartesian coordinates are $x \equiv \phi$ and $z$. All unperturbed quantities depend on the height above the disc $z$ only where $\vec{B}_0 = B_0(z) \vec{e}_\phi$ (see Fig. \ref{F:SPar}).

\begin{figure}
    \centering
    \includegraphics[width=\linewidth]{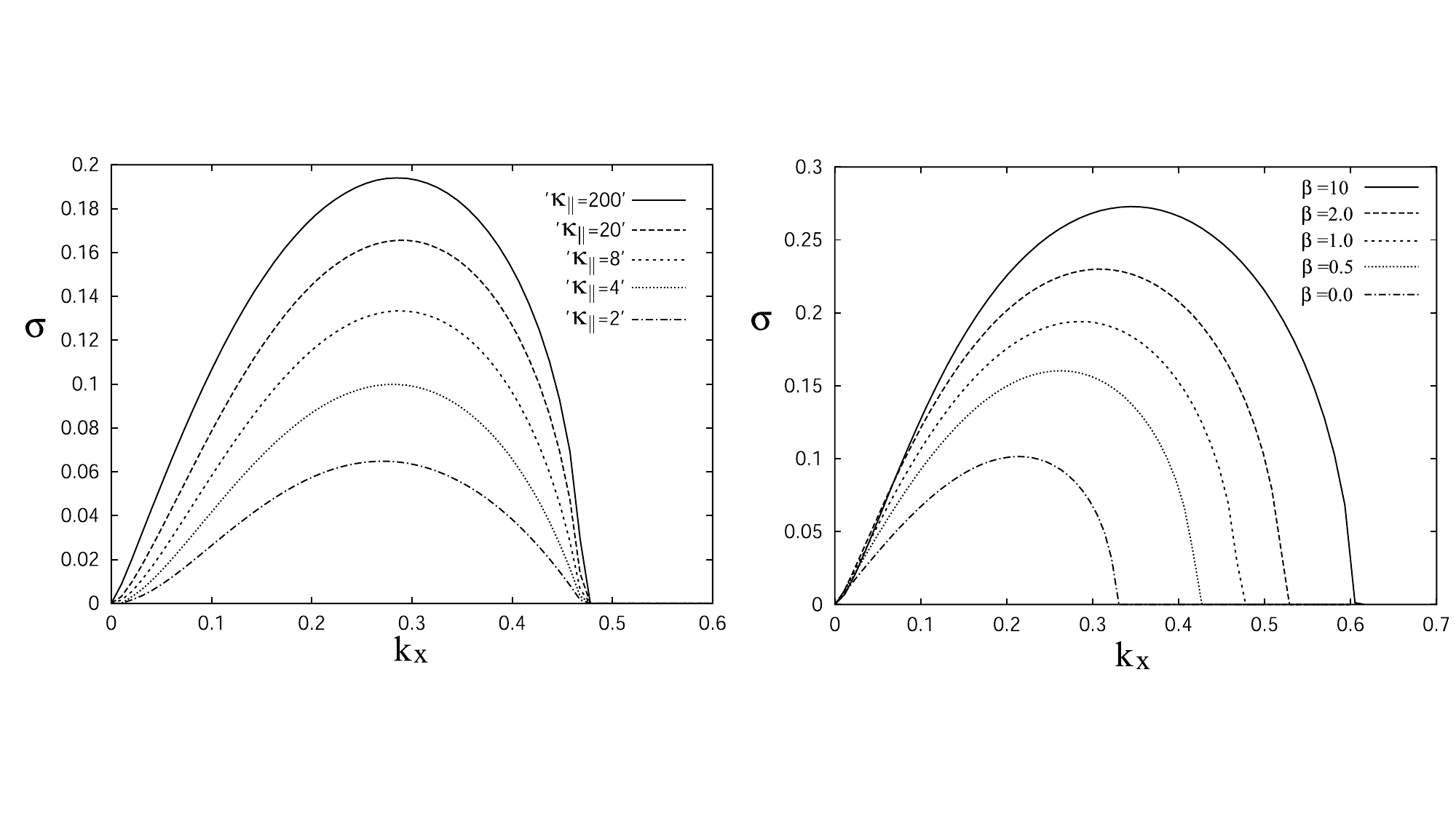}
    \caption{Left: the growth rate parallel to the magnetic field as function of the mode wave number for different values of the parallel diffusion coefficient with $P_{\rm CR} = P_g$. Right: the growth rate parallel to the magnetic field as function of the mode wave number for different values CR to gas pressure ratio with $\kappa= 200$. The other parameters are: $P_g = P_m$ and $\Omega = 0$. By courtesy of\cite{ref:ku04}.}
    \label{F:TK04}
\end{figure}

The equilibrium model satisfies a pressure balance 
\begin{equation}
    {d \over dz} \left(P_g + P_m + P_{\rm CR} \right) + \rho g_{\rm ext} = 0 \ ,
\end{equation}
$g_{\rm ext}$ stems for the effect of an external gravitational field due for instance to stars. They supplement it by a disc-halo model where the gas temperature increases as an hyperbolic tangent with z from the disc ($T=10^4$ K) to the halo ($T=2.5~10^5$ K). The perturbation analysis is conducted with $\vec{k}$ lying in the disc plane, it leads to a system of 2 differential coupled Eqs (see Eq. 10 in \cite{ref:ku04}) which have to be integrated with proper boundary conditions at $z=0$ and $z \rightarrow \infty$. The disc is assumed to be symmetric, this imposes the behaviour of the perturbed quantities at $z=0$. The authors only investigate the growth rate for perturbations along the background magnetic field. They find that the growth rate increases with the CR diffusion coefficient (in units of $H c_s$ \footnote{$H$ is the density scale at $z=0$ it reads $H={c_s^2 \over (g \gamma_g)}$}), and the ratio of CR to gas pressure, see Fig. \ref{F:TK04} but decreases with the Coriolis force (see their figure 4) which compensates over the effect of magnetic buoyancy. The wave number $k_{\rm crit, P}$ beyond which the system is stable only depends on the CR to gas pressure ratio; it increases with it. CR diffusion suppresses any effect of an excess of CR pressure. Only if the diffusion coefficient is reduced to zero the system is strongly stabilised \cite{ref:ryu} because CR pressure opposes the gas motion along the magnetic field and the critical wave number drops (see Eq.\ref{Eq:KCRIT}). Otherwise any, even small, diffusion coefficient allows the instability growth rate to increase as the effect of CR pressure gradient along the magnetic field tends to oppose the gas fall into the footpoints of the magnetic loops.\\ 

Kuwabara et al \cite{ref:ku06} introduce a gravitational potential $\psi_G$ of the self-gravity, hence link to the gas density by a Poisson Eq. $\nabla^2 \psi_G = 4\pi G \rho$, where G is the gravitational constant. This supplementary term allows them to treat the effect of Jeans instability in parallel to magnetic buoyancy effects (the Parker instability). Self-gravitation complicates the analysis substantially, the authors have now a system of 4 differential coupled Eqs (see Eq. 23 in \cite{ref:ku06}). Their analysis is also conducted in 3D which allows them to derive the growth rate perpendicular to the background magnetic field direction. A general trend is that CRs do not have an influence over the wave number $k_{\rm crit}$ marking the stability of the system (the system is stable above this value) as soon as the diffusion coefficient differs from zero. In the case of no magnetic field this critical wave number is the one for the Jeans instability and reads $k_{\rm crit, J}= \sqrt{4\pi G \rho/c_s^2}$. Otherwise, again CRs add a pressure to the gas hence producing a larger sound speed $c_{s,t}= \sqrt{c_s^2 + c_{\rm CR}^2}$ which reduces $k_{\rm crit, J}$. Another general result is the independence of the growth with respect to the CR parallel diffusion coefficient in the case of perturbations perpendicular to the magnetic field ($k_\phi=0$). However, the growth rate for parallel perturbations ($k_r =0$) is sensitive to $\kappa$ as discussed above. Both instabilities compete, this competition is arbitrated by the CR diffusion coefficient and the CR to gas pressure ratio. A strong CR pressure tends to suppress the Jeans instability and favours the Parker instability. CR diffusion decouples the gas from CRs and reduces the effect of CR pressure. 

\section{Main Cosmic-Ray-driven instability classes}\label{S:MICR}
\subsection{Solving for kinetic plasma instabilities}
We describe succinctly here the procedure necessary to derive the dispersion relation from the Vlasov-Maxwell system (Eqs.\ref{Eq:Vlasov}-\ref{Eq:Thomson}). We will apply this procedure to the case of the streaming instability in Sect. \ref{S:LINEM}.\\

The procedure requires to linearise the Vlasov-Maxwell system, so to decompose all quantities in an unperturbed part and a small perturbed correction. For the particle distribution one has $f_a = F_a + \delta f_a$, then $\vec{E}= \delta \vec{E}$ (we assume no background electric fields) and $\vec{B}= \vec{B}_0 + \delta \vec{B}$. The equilibrium state $F_a$ is considered to be known, it verifies the Vlasov-Maxwell system. We can then extract an equation for the first order perturbation $\delta f_a$ by subtracting the equilibrium state Eqs to the original ones.
This reads:
\begin{equation}
    \begin{split}
 \partial_t \delta f_a  + \vec{v}. \vec{\nabla} \delta f_a + & q_a \left( {\vec{v} \over c} \wedge \vec{B}_0\right).\partial_{\vec{p}} \delta f_a = -q_a \left(\delta\vec{E} +{\vec{v} \over c} \wedge \delta  \vec{B}\right).\partial_{\vec{p}} F_a \\
\vec{\nabla}.\delta \vec{E} = & ~4\pi \sum_a q_a \int d^3\vec{p} \delta f_a(\vec{x},\vec{p},t)  \\
\vec{\nabla} \wedge \delta \vec{B} = &{1 \over c} \partial_t \delta \vec{E}+ {4\pi \over c} \sum_a q_a \int d^3\vec{p} \vec{v} \delta f_a \\
\vec{\nabla} \wedge \delta \vec{E} = &-{1 \over c} \partial_t \delta \vec{B} \\
\vec{\nabla}.\delta \vec{B} =& 0 
\end{split}
\end{equation}
This system is the starting point of the linear analysis. 
Krall \& Trivelpiece \cite{ref:kra} (hereafter KT73) discuss different cases: in \S 8.3 the case of electrostatic ($\delta E \neq 0$, $\delta B=0)$ waves with no background magnetic field ($B_0=0$), in \S 8.9 the case of both electrostatic and electromagnetic waves with no background magnetic field. We make a stop on the last case: electrostatic and electromagnetic waves with a background magnetic field treated in \S 8.10. The authors only consider for the equilibrium distribution $F_a$ an anisotropic function $F_a(v_\perp^2, v_\parallel)$. Here we adopt the velocity components parallel and perpendicular to $\vec{B}_0 = B_0 \vec{z}$ and consider the background component to be non-relativistic (as composed by a background thermal plasma in our CR problem). The contribution of CRs to the dispersion relation has to be then added by-hand to this analysis. So from this point we set $\vec{p}= m\vec{v}$ because we first discuss non-relativistic background plasma contribution. One obvious choice is the anisotropic Maxwellian distribution 
\begin{equation}\label{Eq:FA}
    F_a= n_a {m_a \over 2\pi k_{\rm B} T_\perp} \left({m_a \over 2\pi k_{\rm B} T_\parallel}\right)^{1/2} \exp\left(-{m_a \over 2 k_{\rm B}} \left({v_\perp^2 \over T_\perp}+ {(v_\parallel-u_d)^2 \over T_\parallel}\right)\right)\ , 
\end{equation}
where $n_a$ is the particle species a density (notice that we normalised our distribution differently from KT73, since $F_0$ gives the particle density instead of the probability density). We have introduced the possibility to have a drift speed $u_d$ along $\vec{z}$ (we prepare here the calculation to be performed in the frame where the CR distribution is isotropic, hence in that frame the background plasma drifts with a speed $u_d$). KT73 used the same expression but with $u_d=0$ (they derived the dispersion relation of normal modes not the instability growth rate). Then the above system of Eqs is solved using the method of characteristics with the unperturbed particle trajectory, i.e. with the Larmor motion in the background magnetic field. This is a basic assumption of the {\bf quasi-linear theory} beside the low amplitude of perturbed electromagnetic fields. Hence, the growth rate derived below is the linear growth rate. It is strictly valid for a period of time before wave-wave interactions start to invalidate the assumptions and the system to enter in a non-linear growth phase. 

KT73 use a direct integration in configuration space leading to
\begin{equation}
    \delta f_a(\vec{x},\vec{v}, t)= -{q_a \over m_a} \int_{-\infty}^t dt' 
    \left(\delta \vec{E}(\vec{x}',t') + {\vec{v}' \over c} \wedge \vec\delta{B}(\vec{x}',t')\right)_{\rm unper}.\vec{\nabla}_{\vec{v}'}F_a(\vec{x}',\vec{v}') 
\end{equation}
The primes mark the Larmor motion. It is assumed above that the perturbed distribution vanishes as $t \rightarrow -\infty$. The equilibrium function is by definition independent of time along unperturbed particle orbits (see KT73 \S 7.7). Then the electric perturbation is expressed in terms of its Fourier transform $\delta \vec{E}= \vec{E}_k \exp(i\vec{k}.\vec{x}-i\omega t)$, the Faraday equation gives $\vec{B}_k={c \over \omega} \vec{k} \wedge \vec{E}_k$. 
The Fourier term involves components like $\exp (-i k_\perp v_\perp\sin(\phi-\Omega_c t) )= \sum_{n=-\infty}^{+\infty} J_n(W) \exp(-in(\phi-\Omega_c t)$, where $J_n$ is the Bessel function of the first kind of order nand $W= k_\perp v_\perp / \Omega_c$. The final form of the perturbed function is given by Eq. (8.10.8) in K73. The perturbed current can finally be expressed in terms of the perturbed electric field $\delta \vec{J}_k= \sum_a q_a \int d^3\vec{v}~\vec{v} \delta f_{\rm k, a} = \bar{\bar{\sigma}}. \vec{E}_k$, $\bar{\bar{\sigma}}$ is the 3x3 conductivity tensor.\\
The general form of the dispersion relation
\begin{equation}\label{Eq:DIS}
   \left( {\omega \over c}\right)^2 \vec{E}_k + \vec{k}\wedge\vec{k}\wedge \vec{E}_k + i\omega {4\pi \over c^2} \bar{\bar{\sigma}}. \vec{E}_k = 0 \ .
\end{equation}
This dispersion relation can be expressed as $\rm{DET}({\bar{\bar{D}}})=0$, where the 3x3 ${\bar{\bar{D}}}$ matrix is given in Eq 8.10.11 in K73. The fluid theory (based on MHD Eqs only) can be recovered by assuming that $(T_\parallel, T_\perp) \rightarrow 0$ which implies that the Bessel functions can be approximated by $J_n(W) \rightarrow {1 \over n!} (W/2)^n$. The final form of the dispersion relation is 
\begin{eqnarray}\label{Eq:SUS}
    {k^2 c^2 \over \omega^2} - 1 - \sum_s X_s &=& 0 ,~\rm{or} \\
    \omega^2 {V_a^2 \over c^2} \left(1+ \sum_s X_s\right) -k^2 V_a^2 &=& 0 \nonumber 
\end{eqnarray}
where $X_s$ is the so-called susceptibility of species s. We will derive in section \ref{S:CRX} the susceptibility of CRs and in \ref{S:TEM} the susceptibility of the thermal background plasma. Below, in section \ref{S:LINEM} we will mostly concentrate on the study of the triggering of a sub-class of perturbations namely circularly-polarized electromagnetic perturbations propagating along the background magnetic field. Beforehand, we have a qualitative discussion about the different instabilities triggered by CRs, all these necessitate to derive the CR susceptibility which we report here as a function of the CR distribution $F_{\rm CR}$ (again a detailed derivation of $X_{\rm CR}$ is provided in section \ref{S:CRX})
\begin{equation}
    X_{\rm CR}={4\pi q^2 \over \omega} \int_0^{p_{\rm max}} dp \int d\mu {p^2 v(p) (1-\mu^2) \over \omega -kv\mu \pm \Omega_s} \left( \partial_p F_{\rm CR} (p, \mu)+ (\mu+{kv \over \omega}) \partial_\mu F_{\rm CR}(p,\mu) \right) \ .
\end{equation}
In order to obtain the contribution of CRs to the dispersion relation we need to fix a particular distribution $F_{\rm CR}$. The latter depends on the problem under consideration, e.g. monoenergetic, power-law, kappa distribution ... etc. In most of the instability studies mentioned below a power-law is adopted. 

\subsection{An important instability: the Weibel instability}
This instability is not necessarily associated with the presence of CRs, however it has an essential importance in astrophysics because it allows to generate magnetic fields from "nothing" (an unmagnetised medium) but charged anisotropic distributions or relative drifts \cite{ref:wei, ref:fre}. The Weibel instability (WI) is considered as the main instability mediating the formation of collisionless shocks themselves likely sites of CR acceleration. We briefly discuss it below. In the case the instability is induced by the presence of two counter streaming beams is called the current filamentation instability (CFI). The different type of beam-driven instabilities is reviewed by \cite{ref:bre}.\\

The simplest version of the WI
results from the off-equilibrium configuration produced by two beams of cold electrons with the same density but moving at the same speed $v_b = \beta_b c$ in opposite directions. The WI is electromagnetic ($\vec{E} \perp \vec{k}$), transverse ($\vec{E} \perp \vec{B}$), with a transversal propagation ($\vec{k} \perp \vec{v}_b$), the perturbed electric field is along the beam propagation axis. In the cold approximation we can use the continuity and momentum Eqs; to derive the perturbed electron densities and speeds supplemented by the Maxwell Eqs to link the current and the perturbed electric field. If we consider the beams moving along the x axis and the wave vector along the y axis the growth rate is then
\begin{equation}
    \Gamma \simeq \omega_{pe} {v_b \over c} \left(1 + \left({\omega_{pe} \over k_yc}\right)^2 \right)^{-1/2}\ .
\end{equation}
It can be seen that at short wave lengths ($k_y \rightarrow 0$) the growth rate is $\Gamma \simeq v_b k_y$, whereas at long wavelength ($k_y \rightarrow \infty$) $\Gamma \simeq \omega_{pe} \beta_b$. In the latter case temperature effects can hamper the instability growth. If the beams are relativistic with a Lorentz factor $\gamma_b=(1-\beta_b^2)^{-1/2}$ the growth rate is now
\begin{equation}
    \Gamma \simeq \omega_{pe} {\beta_b \over \sqrt{\gamma_b}} \left( 1 + \left({\omega_{pe} \over \gamma_b^{3/2} k_yc}\right)^2  \right)^{-1/2}\ .
\end{equation}
The relativistic fluid theory in case of an electron-ion plasma is discussed in e.g. \cite{ref:gra}, for the kinetic theory see \cite{ref:ach}. The case of an electron-positron plasma is treated in e.g. \cite{ref:yang}.

\subsection{Main anisotropy frameworks to investigated Cosmic Ray-driven instabilities}
CR driven instabilities involve anisotropic particle distribution with control parameters for the degree of anisotropy with respect to a reference isotropic distribution. Hence, usually one can use a perturbed analysis with a development in anisotropy of the particle distribution. In the case the anisotropy is strong, as in the case of CRs at shocks one can still conduct the calculation in the shock restframe and consider the particle distribution perturbed around its isotropic part. This implicitly assumes that some efficient scattering process ensures quasi-isotropy. This assumption is not always correct, in particular for the highest energy particles escaping the shock structure because of their ballistic motion. In that case, the unperturbed function carries itself some anisotropy and the perturbative approach has to be carried due to the effect of the microscopic electromagnetic fluctuations over the particle distribution function. The distribution function is developed in anisotropy over the particle pitch-angle, but the distribution is assumed to be gyrotropic. This means that the Larmor time is short compared to scattering time. We have 
$F(p,\mu)={n_{\rm CR} N(p) \over 4\pi} \left(1 + 3 {u_d \over c}\mu + {\chi \over 2} (3\mu^2-1)+ ...\right)$, where $\mu = \cos(\vec{v}, \vec{B})$ and $u_d$ and $\chi$ are two parameters controlling the anisotropy. A more general formalism gives 
\begin{equation}\label{Eq:F2}
F(p,\mu)= f_0(p) + {\vec{p} \over p}.\vec{f}_1(p) + {5 \over 2} {\vec{p}\vec{p} \over p^2}:\bar{\bar{f}}_2(p) + ...,
\end{equation}
the tensor $\bar{\bar{f}}_2$ is linked to the CR pressure by
\[
\bar{\bar{P}}_{\rm CR} = \int d^3\vec{p} \vec{p}\vec{v} f(p) \simeq {4\pi \over 3} \int d p \left(\bar{\bar{I}} f_0 + \bar{\bar{f}}_2(p)\right) \ .
\] 
A general approach to these problems exists, it uses a Vlasov-Boltzmann approach including a spherical harmonic (SH) expansion of the particle distribution \cite{ref:tzo, ref:tzo2, ref:bel6}. We briefly discuss this approach below as it permits to investigate instabilities triggered by high order terms in the SH developments using a general formalism. The SH Vlasov approach provides a general formalism to write the CR distribution \cite{ref:tzo2} without assuming gyrotropy. The particle distribution function reads 
\begin{equation}\label{Eq:FSH}
    F(\vec{r},\vec{p},t)= \sum_{\ell=0}^\infty \sum_{m=-\ell}^{m=+\ell} f_\ell^m(\vec{r},p,t) P^{|m|}_\ell(\mu) \exp(im\phi) \ ,
\end{equation}
with $f_\ell^{-m}(\vec{r},p,t)=f_\ell^m(\vec{r},p,t)^*$ (its complex conjugate) and $P^{|m|}_\ell(\mu)$ is the associated Legendre polynomial of degree $m$ order $\ell$, they are deduced from the ordinary Legendre polynomials using
\[
P_\ell^m(x) = (-1)^m (1-x^2)^{m/2} {d^m \over d x^m} P_\ell(x) \ ,
\]
Finally, $\phi$ is the azimuthal angle associated with the gyromotion. The previous expansion has to be truncated at some fixed values for $\ell$ and $m$. Eq. \ref{Eq:FSH} can be inserted into the Boltzmann Eq. 
\begin{equation}\label{Eq:BOL}
    \partial_t F(\vec{x},\vec{p},t) + \vec{v}. \vec{\nabla} F + q \left(\vec{E} + {\vec{v} \over c} \wedge \vec{B}\right).\partial_{\vec{p}} F = \partial_t F_c \ ,
\end{equation}
where RHS is controlled by collisions. This leads to Eq.16 in \cite{ref:tzo2} giving the time evolution of 
$f_\ell^m$:
\begin{equation}
    \partial_t f_\ell^m= A + E + B + C + S \ ,
\end{equation}
where A are advection terms, E electric terms, B magnetic terms and C collision terms, finally S is a source term. Each terms A, E, B, C are lengthy and will not be reported here (see \cite{ref:tzo2} for further details).

\subsection{Instabilities due to pressure anisotropy: the mirror/firehose instabilities}
\begin{figure}
    \centering
    \includegraphics[width=\linewidth]{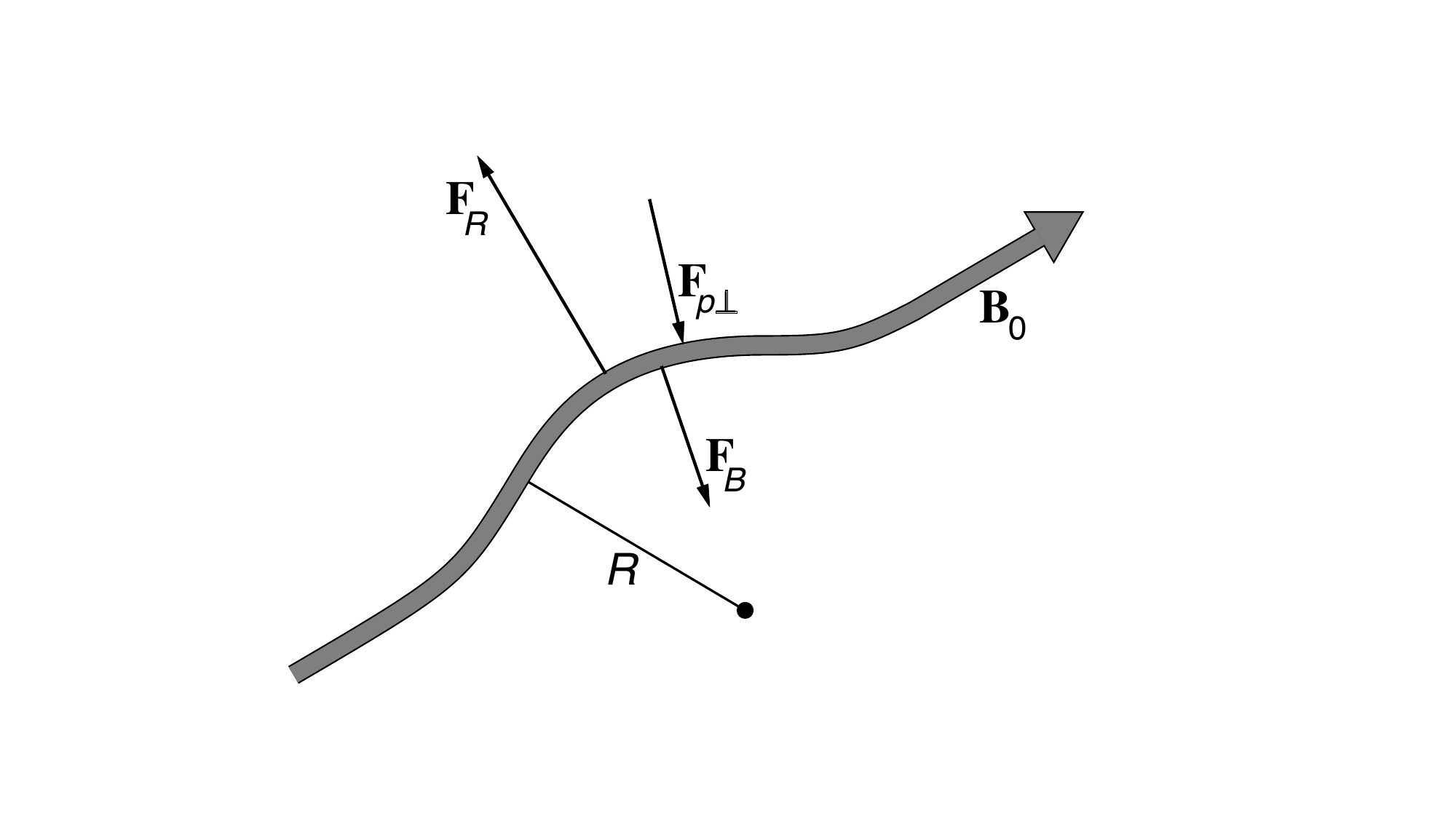}
    \caption{Mechanism of the firehose instability. By courtesy \cite{ref:tre} (Imperial College press 1997).}
    \label{F:FIR}
\end{figure}

These instabilities are triggered because of an excess of pressure either along the background magnetic field (firehose) $P_\parallel$ or perpendicularly $P_\perp$ (mirror) to the background magnetic field. These instabilities do not explicitly need the presence of CRs to be triggered, they can be driven by anisotropic plasma temperatures. Kinetic calculations lead to the triggering criterion \cite{ref:has}, noting $\beta=8\pi n m V_T^2/B^2$, with $V_T=\sqrt{k_B T/m}$ the thermal speed. In the firehose case we have:
\begin{equation}\label{Eq:FCRI}
   \beta_\parallel - \beta_\perp > 2 \ ,
\end{equation}
In the mirror case we have:
\begin{equation}\label{Eq:FCRI}
  \left( \beta_\perp - \beta_\parallel \right) {\beta_\perp \over \beta_\parallel}> 1 \ .
\end{equation}
The origin of the firehose instability is clearly described in \cite{ref:par58} based on a magneto-hydrodynamic treatment differentiating parallel and perpendicular gas motion and pressure with respect to the background magnetic field. The instability is due to the competition between the plasma flow along a magnetic flux tube producing a centrifugal force $F_R=\rho u_\parallel^2/R$, where $R$ is the curvature radius of the flux tube and restoring forces: perpendicular thermal pressure $F_{p_\perp}=|\vec{\nabla} P_\perp| \simeq {P_\perp \over R}$ and magnetic stress $F_B= {B^2 \over 4\pi R}$ (see figure \ref{F:FIR}). As explained by \cite{ref:par58} the instability results from the centrifugal force produced by a train of beads moving along the magnetic field which amplifies a transversal motion. The instability growth rate is maximal for parallel propagation ($k=k_\parallel$) and vanishes in case of perpendicular propagation. The instability develops at $k < k_0(P_\parallel, P_\perp)$ ($k_0$ is given in Eq. 3.82 of \cite{ref:tre}). 

The mirror instability develops for perpendicular modes ($k=k_\perp$). It requires a kinetic treatment to account for particle trapping in magnetic mirrors. Particles trapped in the mirror get accelerate towards lower magnetic field regions by the $-\mu_d \nabla B$ force ($\mu_d =1/2 mv_\perp^2/B$ is the particle magnetic moment), this diamagnetic repulsion enhanced the perpendicular pressure and excludes the magnetic field further on leading to an instability. \\

The firehose/mirror instability including CRs has been discussed at different levels. First, works only consider the effect of CRs over the triggering process, put in other words these works investigate relativistic-pressure-driven instabilities see \cite{ref:wen, ref:noe, ref:sco, ref:osi}. Another calculation given by \cite{ref:ach13}, proposes an estimation of the impact of the presence of CRs with a finite current over these class of instabilities. Finally Bykov et al \cite{ref:byk13} propose an analysis of the combined effects of CR streaming and pressure anisotropies. Below we discuss the work by \cite{ref:osi}.\\

In order to capture the pressure-driven instabilities {\it only}, the mean CR distribution function has to have a quadrupole anisotropy, namely 
\begin{equation}
    F(p,\mu)={n_{\rm CR} N(p) \over 4\pi} \left(1 + {\chi \over 2} (3\mu^2-1) \right) \ ,~\rm{with}~\chi < 1,
\end{equation}
and $\int dp p^2 N(p) = 1$. The parallel and perpendicular CR pressures are 
$P_\parallel =\int d^3{\vec{p}} pv \mu F(p,\mu)$ and $P_\perp= \int d^3{\vec{p}} pv (1-\mu^2) \cos\phi^2 F(p,\mu)$ ($\phi$ is the gyrophase). We note $\Delta P=P_\parallel-P_\perp$. Solving to first order perturbation in the MHD and Vlasov Eqs. (see the full treatment in section \ref{S:LINEM} for the case of the streaming instability) one gets the following dispersion relation
\begin{eqnarray}
    \left(\omega^4-\omega^2(c_s^2+ V_a^2)k^2 + V_a^2c_S^2k^2k_\parallel^2
    +\left(\omega^2-k_\parallel^2c_s^2\right) {\Delta P \over \rho}(k_\parallel^2-2k_\perp^2)\right) && \nonumber \\
  \times \left(\omega^2-k^2_\parallel V_a^2 + {\Delta P \over \rho} k_\parallel^2\right)  = 0 && \ .
\end{eqnarray}
This combines both oblique mirror ($k_\perp \neq 0$) and firehose ($k_\perp =0$) cases \cite{ref:osi}. The instabilities only develop for long wave length, i.e. $kv \ll \Omega_s$ and $\omega \ll \Omega_s$, where $\Omega_s$ is the relativistic particles gyro-frequency.\\

Let us consider first the case with $k_\perp =0$, the dispersion relation becomes
\begin{equation}
   \left(\omega^2-k^2 c_s^2 \right)\times \left(\omega^2-k^2V_a^2+ {\Delta P \over \rho} k^2\right) = 0.
\end{equation}
The system is unstable if $\Delta P >0$ and  ${\Delta P \over \rho} > V_a^2$ (the other branch being the sonic mode). The growth rate is 
\begin{equation}
    \Gamma=k \sqrt{{\Delta P \over \rho}-V_a^2} \ .
\end{equation}
The mirror case is obtained for $k_\parallel =0$, we get
\begin{equation}
 \omega^2-k^2 (V_a^2+ c_s^2 + 2 {\Delta P \over \rho}) = 0 .
\end{equation}
The system is unstable if $\Delta P  < 0$ and $2 {|\Delta P| \over \rho} > (V_a^2+ c_s^2)$.
The growth rate is
\begin{equation}
    \Gamma=k \sqrt{2{|\Delta P| \over \rho}-V_a^2-c_s^2} \ .
\end{equation}

\subsection{The acoustic or Drury instability}
This instability is triggered because of the presence of a strong CR pressure gradient, as it is the case in CR mediated-shock precursors. This force induces the destabilisation of sonic perturbations \cite{ref:dru84}. The origin of the instability stems on the effect of the force associated to the CR pressure gradient in the precursor over small gas density fluctuations. Unless in a very special case where CR pressure gradient fluctuations are exactly proportional to density fluctuations the force exerted is non uniform leading to gas velocity fluctuations which themselves amplify the original density fluctuations. This is the essence of the acoustic or Drury instability. Let us put this in a more formal way following the analysis by \cite{ref:dru86,ref:kan92}. The main criterion for the instability to be triggered is :
\begin{equation}
    1 < k {P_{\rm CR} \over |\vec{\nabla} P_{\rm CR}|} < k (1+\beta_\kappa) {\bar{\kappa} \over \gamma_{\rm CR} c_s} \ ,
\end{equation}
where $\beta_\kappa = {\partial \ln(\bar{\kappa}) \over \partial \ln(\rho)}$. We will hereafter note $L_{\rm CR} = \left({P_{\rm CR} \over |\vec{\nabla} P_{\rm CR}|}\right)$, the CR shock precursor size. This criterion can be rewritten as $M_s > {\gamma_{\rm CR} \over (1+\beta_\kappa)} {L_{\rm CR} u_{\rm sh} \over \bar{\kappa}}$, where $M_s={u_{\rm sh} \over c_s}$ is the shock sonic Mach number. The averaged parallel diffusion coefficient is $\bar{\kappa} = \int d^3 p p \nabla f \kappa(p) /\int d^3 p p \nabla f $, the gradient is applied along the shock normal.\\
The linear analysis is applied to a system \ref{Eq:CRMHD1}-\ref{Eq:CRMHD4} but dropping the magnetic terms, with the CR flux being estimated as $\vec{F}_{\rm CR} =-\bar{\kappa} \vec{\nabla} e_{\rm CR}$. The unperturbed quantities are defined by the CR shock precursor of size $L_{\rm CR}$, hence gas density and pressure and CR pressure varies as $\exp(x/L_{\rm CR})$ where $x$ is the path along the shock normal. The linear analysis yields to the following dispersion relation

\begin{eqnarray}
    \omega^3 + i\bar{\kappa}{\omega^2 \over L_{\rm CR}^2}\left(1-ikL_{\rm CR}\right)^2 - i{k\omega \over L_{\rm CR}}\left(1-ikL_{\rm CR}\right) \left({{\gamma_g} P_g \over \rho}+ {\gamma_{\rm CR} P_{\rm CR} \over \rho}\right) && \nonumber \\
    -\left(1-ikL_{\rm CR}\right)^3 \left(i{P_{\rm CR} \bar{\kappa} (1+\beta_\kappa) \over L_{\rm CR}^4 \rho} - {\gamma_g P_g \over \rho L_{\rm CR}^4} \right) = 0 && \end{eqnarray}

\begin{figure}
    \centering
    \includegraphics[width=1.1\linewidth]{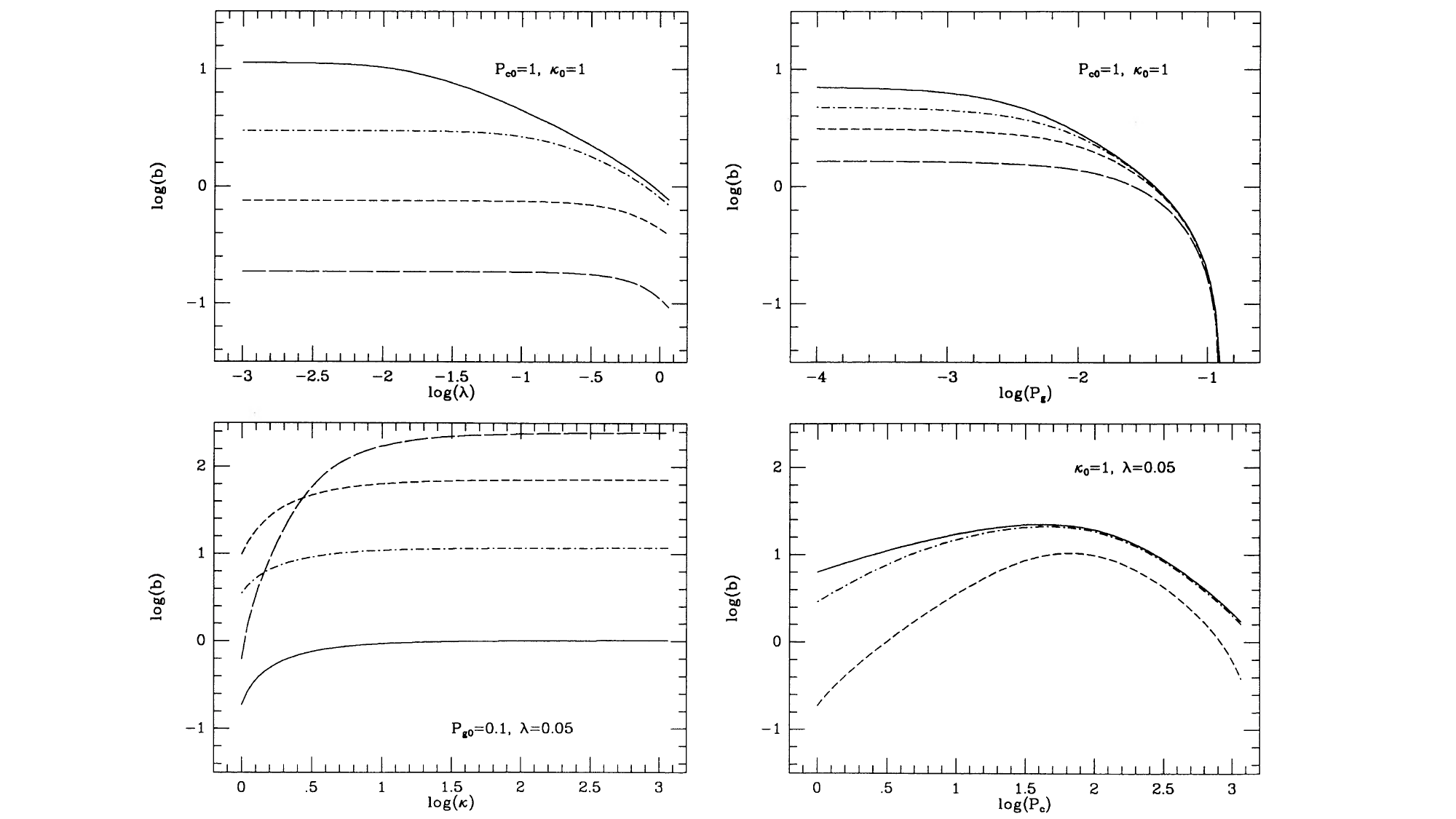}
    \caption{Acoustic instability growth rate as a function of different quantities: upper-left the wavelength (solid line $P_g=10^{-4}$, dot-dashed line $P_g=10^{-3}$, dashed line $P_g=0.01$, long dashed line $P_g=0.1$), upper-right: the gas pressure (solid line $\lambda=0.05$, dot-dashed line $\lambda=0.1$, dashed line $\lambda=0.2$, long-dashed line $\lambda=0.5$), lower-left: CR mean diffusion coefficient (solid line $P_{\rm CR}=1$, dot-dashed line $P_{\rm CR}=10$, dashed line $P_{\rm CR}=100$, long dashed line $P_{\rm CR}=1000$), lower-right: CR pressure (solid line $P_g=10^{-3}$, dot-dashed line $P_g=10^{-2}$, dashed line $P_g=0.1$). In all plots $L_{\rm CR}=1, \rho=1, \gamma_{\rm CR}=\gamma_g=5/3$. By courtesy of  \cite{ref:kan92}.}
    \label{F:DRU}
\end{figure}

The growth rate is plotted in Fig. \ref{F:DRU} as function of different normalised quantities entering in the problem. It shows that the growth rate is weakly dependent on the wavelength but restricted to small scale fluctuations. Lower initial gas pressures lead to a higher growth rate as the impact of the CR pressure gradient is enhanced. The stronger the CR pressure the higher the growth rate (lower-left). Low diffusivity and high CR pressure tend to reduce the growth rate because of the friction effect \cite{ref:ptu} (lower-right and lower-left). Zank et al \cite{ref:zan} derived the dispersion relation including a magnetic field of different orientation with respect to the shock normal. The magnetic field amplification (and somehow saturation) is discussed in \cite{ref:dow12} and \cite{ref:dow14} using 2 and 3D MHD simulations of fast shocks. It is shown that the instability is efficient at amplifying a background magnetic field by at least one order of magnitude.\\

The acoustic instability is not limited to CR shock precursors but can be adapted to interstellar medium contexts \cite{ref:beg, ref:tsu} including the effect of CR streaming (see next section). It is shown that this instability can lead to the generation of multiple shocks.

\subsection{The streaming instability and its relatives: the filamentation and oblique mode instabilities}
The streaming instability is discussed in details in section \ref{S:LINEM}. This instability is triggered because of a drift motion of the CR population with respect to the background gas and/or because CR have an anisotropic distribution (dipole-like, with $F(p,\mu)={n_{\rm CR} N(p) \over 4\pi} \left(1 + 3 {u_d \over c}\mu \right)$), see \cite{ref:byk13}. The origin of the drift is multiple fold. In particular, if CRs are accelerated at shock waves, their remain confine in a small area upstream the shock front (the CR precursor). This precursor as seen from in the interstellar gas restframe resembles to a charged layer moving at a speed $u_{\rm sh}$ which is almost the drift speed for fast super-Alfv\'enic shocks. CRs carry then a strong source of free energy through their streaming which is able to trigger two branches of the instability either the resonant branch or the non-resonant branch. The streaming instability creates magnetic fluctuations which can be at the origin of the turbulence in CR mediated shocks.  

\subsubsection{Instabilities associated with a CR drift}
The recent literature has discussed some CR-driven instabilities that build up on top of the triggering of the streaming instability at shock fronts. These are 1) the filamentation instability \cite{ref:rev12} and 2) the oblique-mode instability \cite{ref:byk11}.\\

The filamentation instability \cite{ref:rev12} results from the growth of magnetic loops at the shock precursor in the non-linear growth phase of the non-resonant streaming instability \cite{ref:bell05}. The origin of the instability is due to the expansion of magnetic loops with a preferential polarisation by the return current compensating the CR current of a given species (mostly protons), whereas the oppositely polarised loops drop. Magnetic loops will concentrate CR in filaments which locally increase the CR current and the source of instability. Reville \& Bell \cite{ref:rev12} derive the instability growth rate as (for a shock speed $u_{\rm sh}$)

\begin{equation}
    \Gamma = \eta \left(u_{\rm sh} \over c\right)^2 \left({U_{\rm CR} \over \rho u_{\rm sh}^2}\right) {e \delta B_{\rm NR} \over m c \gamma_{\rm min}}
\end{equation}
$\delta B_{\rm NR}$ is the amplitude of the magnetic field produced by the non-resonant 
streaming instability. $\gamma_{\rm min} mc$ is the minimum particle momentum contributing to magnetic amplification at some distance. $U_{\rm CR}$ is the CR energy density, $\eta$ depends on the particle distribution, eg for a $p^{-4}$ distribution it is $1/\sqrt{p_{\rm max}/p_{\rm min}}$.\\

\begin{figure}
    \centering
    \includegraphics[width=1.1\linewidth]{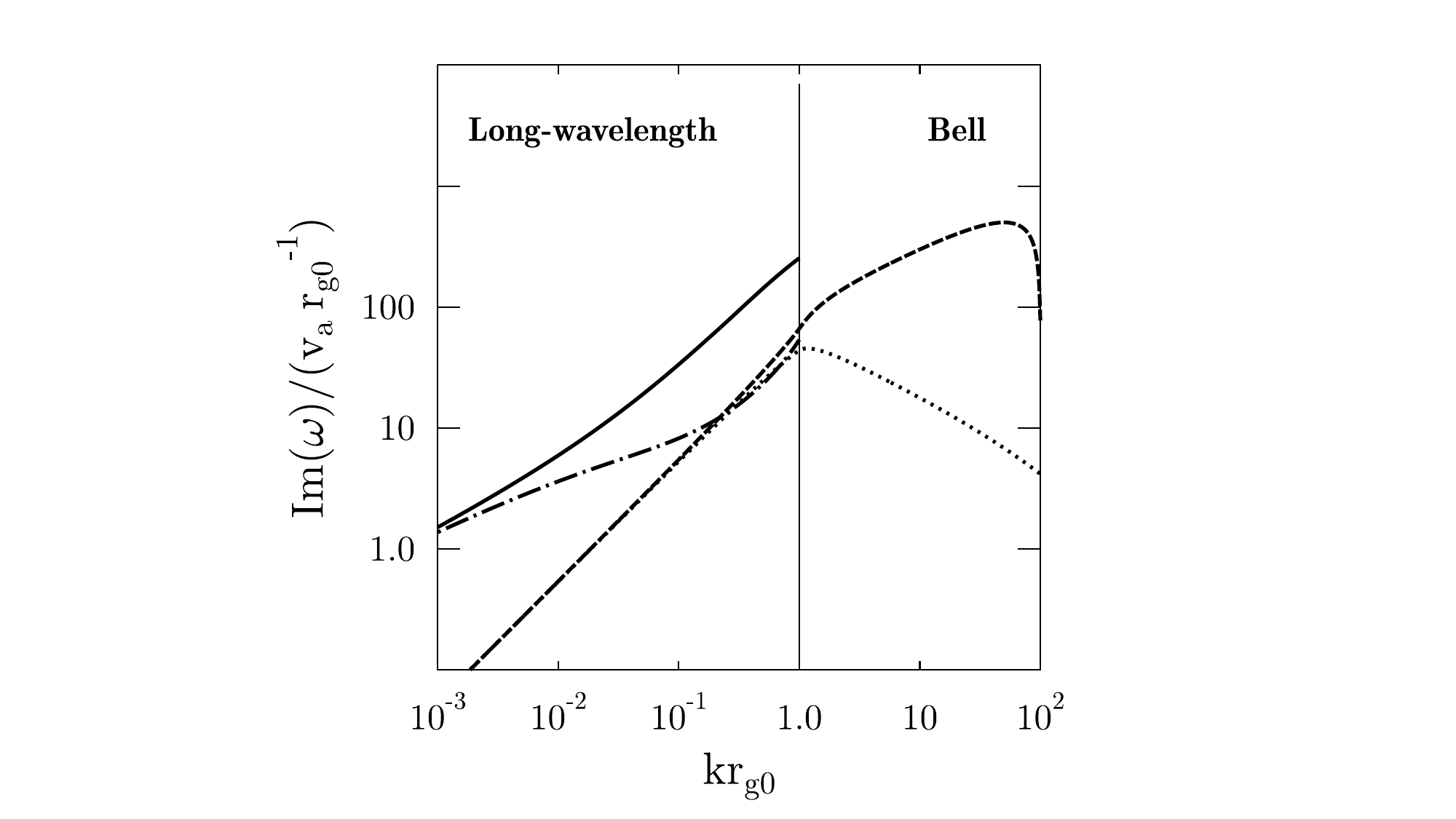}
    \caption{Growth rate of long wavelength modes. Solid and dashed lines show the behaviour of two cricular polarisations of oblique long wavelength modes. In comparison dashed and dotted curves show the solution in case just streaming modes are present. $\eta=10$ is assumed. By courtesy of \cite{ref:byk11}.}
    \label{F:BykovLong}
\end{figure}

The oblique mode instability \cite{ref:byk11} results from a dynamo process initiated by the magnetic field seeds produced by the non-resonant streaming instability. The calculation is done in several steps. First, the equations controlling the velocity and magnetic amplitudes of fluctuations produced by the short-scale non-resonant instability are derived. Then, MHD and Vlasov Eqs are averaged over these short-scale fluctuations using a mean field approach used in dynamo theory. The electromotive force $\langle \vec{u} \wedge \vec{b} \rangle$ of the curl products of the velocity and magnetic fields imposed by the non-resonant instability drives the amplification of long-wavelength modes. Ponderomotive forces ensure the momentum transfer between the background plasma and the non-resonant streaming induced turbulence. The maximum growth rate in the long wavelength domain that is for $k \le \lambda_{\rm mfp}^{-1}$, where $\lambda_{\rm mfp}$ is the mean free path of particles with Larmor radius $r_{g}$ is :
\begin{equation}
    \Gamma= \sqrt{{\pi \over 2} \sqrt{{\langle \delta B_{\rm NR} \rangle}^2 \over B_0}} \sqrt{k k_0 \over \eta} V_a \ .
\end{equation}
With $k_0 = 4\pi {J_{\rm CR} \over B_0 c}$, $\eta = \lambda_{\rm mfp}/r_g$. Figure \ref{F:BykovLong} shows the dispersion relation of the long wave modes in the range $k r_g < 1$. It is clear that the instability provides growth rates in the long wavelength range stronger than in the case pure streaming modes are set. 

\section{The Cosmic-Ray streaming instability}\label{S:LINEM}
In this section we present in more details the derivation of the growth rate of the CR-driven streaming instability. 

\subsection{A detailed derivation of the streaming instability growth rate}\label{S:CRX}
For this derivation we consider simple initial conditions: a {\it cold}, {\it static} medium with {\it uniform} density $\rho$ composed of electrons and ions (mostly protons) immersed in an {\it uniform} background magnetic field $B_0$ along the z direction. Then, the Alfv\'en speed is $V_{\rm a}= B_0/\sqrt{4\pi \rho}$. We consider a population of CRs possibly composed of various components termed as $a$ which can be either ions, electrons or positrons.\\

Below we propose a derivation based on a private communication by X.N. Bai.

\subsubsection{Eqs of magnetohydrodynamic linearisation:} We start with the linearisation of the MHD equations in the Newtonian limit. The one-fluid MHD Eqs. are given in Eqs \ref{Eq:RHO}-\ref{Eq:OHM}. CRs contribute to two terms in the Laplace equation entering in RHS of the momentum equation above: 1) the slight difference between background electron $n_e$ and proton $n_p$ density is the CR density $n_{\rm CR} = n_e - n_p$ \footnote{The background plasma is assumed to be composed of a mixture of electrons and protons, while CR are composed of protons. The generalisation to more complex composition is straightforward.} which produces an electric force $e n_{\rm CR} \vec{E}$, 2) the Amp\`ere equation includes a current associated to CRs $\vec{J}_{\rm CR} = e n_{\rm CR} \vec{u}_{\rm CR}$ where $\vec{u}_{\rm CR}$ is the mean CR velocity in the background plasma frame which contributes to the $\vec{J} \wedge \vec{B}$ term. As a further matter of simplification we restrict our analysis to the {\it incompressible} case below, hence $\vec{\nabla}. \vec{u}=0$ and 
\begin{eqnarray}\label{EQ:CRMHDs}
\partial \vec{u} + \vec{u}.\vec{\nabla} \vec{u} &=& {1\over 4\pi \rho} (\vec{\nabla}\wedge \vec{B}) \wedge \vec{B}-{1 \over \rho c} \vec{J}_{\rm CR} \wedge \vec{B} - \vec{\nabla} {P \over \rho} -{e n_{\rm CR} \over \rho} \vec{E}\ , \nonumber \\
{1 \over c} \partial_t \vec{B} &=& -\vec{\nabla} \wedge \vec{E} \ . \nonumber \\
\vec{E} &=& -{\vec{u} \over c} \wedge \vec{B}
\end{eqnarray}

We linearise Eqs \ref{EQ:CRMHDs} by setting 
$\vec{J}_{\rm CR}= \vec{J}_{\rm CR, 0} + \delta \vec{J}_{\rm CR}$, $\vec{B}=\vec{B}_0 + \delta \vec{B}$, $\vec{u} = \delta \vec{u}$ (we assume no background mean fluid motion only a response to the CR pervading effect). The pressure term is relaxed since the perturbations under consideration are transverse \cite{ref:bell04}. This does not however preclude the effect of some non-linear gradient pressure effects over the instability growth. We have:
\begin{eqnarray}\label{EQ:CRMHDl}
\partial_t \vec{u} &=& {1\over 4\pi \rho} (\vec{\nabla}\wedge \delta \vec{B}) \wedge \vec{B}_0-{1 \over \rho c} \vec{J}_{\rm CR,0} \wedge \delta \vec{B} -{1 \over \rho c} \delta \vec{J}_{\rm CR} \wedge \vec{B}_0 -{e n_{\rm CR} \over \rho} \vec{E}\ , \nonumber \\
{1 \over c} \partial_t \delta \vec{B} &=& -\vec{\nabla} \wedge \vec{E} \ . \nonumber \\
\vec{E} &=& -{\vec{u} \over c} \wedge \vec{B}_0 \ .
\end{eqnarray}
In order to proceed we consider {\it parallel propagation perturbations} scaling as $\exp i(kz-\omega t)$. The system \ref{EQ:CRMHDl} reads as :
\begin{eqnarray}
-i\omega \vec{u} &=& {i\over 4\pi \rho} (\vec{k} \wedge \delta \vec{B}) \wedge \vec{B}_0 -{1 \over \rho c} \vec{J}_{\rm CR,0} \wedge \delta \vec{B} -{1 \over \rho c} \delta \vec{J}_{\rm CR} \wedge \vec{B}_0 -{e n_{\rm CR} \over \rho} \vec{E}  \nonumber \ , \\
\omega \delta \vec{B} &=& c\vec{k}\wedge \vec{E} \ .
\end{eqnarray}
We consider {\it electromagnetic perturbations}, we take $\vec{k}.\vec{E} = 0$ \footnote{The assumption of considering electromagnetic perturbations is consistent with the assumption of incompressibility as no electric field develop parallel to the background magnetic field no density perturbations are induced.} and we have $\vec{E}= {\omega \over kc} \delta \vec{B} \wedge \vec{k}/k = {\omega \over kc} \delta \vec{B} \wedge \vec{z}$. We substitute $\delta \vec{B}$ as function of $\vec{E}$ in the momentum Eq, to find (we take the curl product of the previous Eq. with $\vec{B}_0$):
\begin{equation}\label{Eq:ED}
    i\omega \vec{E}= {i k^2 V_{\rm a}^2 \over \omega} \vec{E} -{J_{\rm CR,0 \parallel} B_0 \over \rho c \omega} \vec{k} \wedge \vec{E} +{4\pi V_{\rm a}^2 \over c^2} \delta \vec{J}_{\rm CR} -{e n_{\rm CR} \over \rho c} \vec{E} \wedge \vec{B}_0
\end{equation}
where $J_{\rm CR,0 \parallel}= \vec{J}_{\rm CR,0}. \vec{z}$. We can further combine the second and fourth terms in the RHS part of Eq. \ref{Eq:ED} as
$-{e B_0 n_{\rm CR} \over \rho c} \left(1-{ku_{\rm CR, \parallel} \over \omega}\right)\vec{E}\wedge \vec{z}$.\\

We now define the perturbation polarisation from their electric field by writing $\vec{E} = E(\vec{x} \pm i \vec{y})$ (+ stands for right-handed modes, - stands for left -handed modes): right (left)-handed modes forward propagating along $\vec{z}$ (with $k > 0$) rotate clockwise (counter-clockwise) for an observer looking in the +z direction. Backward modes ($ k < 0$) have the reverse polarisation.\\

As $\vec{E} \wedge \vec{z}=\pm i \vec{E}$ we find:
\begin{equation}\label{Eq:ED2}
    i\omega \vec{E}= {i k^2 V_{\rm a}^2 \over \omega} \vec{E} +{4\pi V_{\rm a}^2 \over c^2} \delta \vec{J}_{\rm CR}^{\pm} \mp i{e B_0 n_{\rm CR} \over \rho c} \left(1-{ku_{\rm CR, \parallel} \over \omega}\right)\vec{E} \ .
\end{equation}

One can readily interpret the different RHS terms in Eq. \ref{Eq:ED2} . The first term leads to the standard dispersion relation of Alfv\'en waves, the second is at the origin of the resonant branch of the streaming instability while the last one is connected to the non-resonant branch and has the threshold term $\left(1-{ku_{\rm CR, \parallel} \over \omega}\right)$ over the parallel CR drift speed.


\subsubsection{Vlasov Eq linearisation}\label{S:VLAL}
The second step is the linearisation of the Vlasov equation, the kinetic equation describing a collisionless plasma and the CR dynamics. The Vlasov Eq. in general reads as
 \begin{equation}\label{Eq:VLA0}
    \partial_t f_a (\vec{x},\vec{p},t) + {c\vec{p} \over \sqrt{m^2c^2+p^2}} . \vec{\nabla} f_a + q_a \left(\vec{E} + {\vec{p} \over \sqrt{m^2c^2+p^2}}  \wedge \vec{B}\right).\partial_{\vec{p}} f_a = 0 \ ,
\end{equation} 
If the species a are assumed to be in the relativistic regime we have: 
\begin{equation}\label{Eq:VLA}
    \partial_t f_a (\vec{x},\vec{p},t) + \vec{v}. \vec{\nabla} f_a + q_a \left(\vec{E} + {\vec{v} \over c} \wedge \vec{B}\right).\partial_{\vec{p}} f_a = 0 \ ,
\end{equation} 
where we have considered the particle velocity and momentum as $\vec{v}$ and $\vec{p}= \gamma m_a \vec{v}$. The particle mass and charge are $m_a$ and $q_a$ respectively. The particle distribution function $f_a$ is evaluated at a position $\vec{x}$, a momentum $\vec{p}$ and a time t.\\

We now consider a stationary and homogeneous \footnote{in fact slowly time and space-dependent with respect to the rapid perturbation response.} Cosmic Ray background distribution $F(\vec{p})$ and a response to the perturbation growth/damping $\delta f(\vec{x}, \vec{p},t)$. So, for each species we have $f_{a}(\vec{x}, \vec{p},t)= F_a(\vec{p})+ \delta f_a(\vec{x}, \vec{p},t)$ (but we drop the index a unless it is necessary). As we restrict this analysis to circularly-polarised parallel propagating perturbations, the perturbative part of the CR distribution function scales as $\exp(i(kz-\omega t))$. It reads 
\begin{equation}
    -i\omega \delta f + i k v_\parallel \delta f+ {q \over c} \left(\vec{v} \wedge \vec{B}_0\right).\partial_{\vec{p}} \delta f = -q \left(\vec{E} + {1 \over \omega} \vec{v} \wedge (\vec{k} \wedge \vec{E}) \right).\partial_{\vec{p}} F
\end{equation}
The aim of this section is to express $\delta f$ in terms of F and reinsert it to derive $\delta \vec{J}_{\rm CR}$ in Eq. \ref{Eq:ED2}.\\

We turn now to work in cylindrical coordinates in momentum, hence the gradient of the "f" terms has the components ($\partial_{p_\perp}, {1 \over p_\perp}\partial_{\phi}, \partial_{p_\parallel}$).
We have
\begin{equation}
   {q \over c} \left(\vec{v} \wedge \vec{B}_0\right).\partial_{\vec{p}} \delta f =-{q v_\perp B_0 \over p_\perp c} \partial_{\phi} \delta f(p_\parallel, p_\perp, \phi) \ . 
\end{equation}
The background CR distribution has a gyrotropic distribution and hence does not depends on $\phi$. We have 
\begin{equation}
    -q \left(\vec{E} + {1 \over \omega} \vec{v} \wedge (\vec{k} \wedge \vec{E}) \right).\partial_{\vec{p}} F=-q\left({kv_\perp \over \omega} \partial_{p_\parallel} F + (1-{k v_\parallel \over \omega})\partial_{p_\perp} F\right) \vec{E}.{\vec{v}_\perp \over v_\perp} \ .
\end{equation}
 We further have 
 \[\vec{E}.\vec{v}_\perp/v_\perp= E_{\rm x} \cos\phi + E_{\rm y} \sin\phi = {1 \over 2} \left((E_{\rm x}-iE_{\rm y})\exp(i\phi)+ (E_{\rm x}+iE_{\rm y})\exp(-i\phi)\right).\] Noting the gyro-frequency $\Omega = qB_0/\gamma mc$, the Eq. reads
 
 \begin{equation}
      -i\omega \delta f + i k v_\parallel \delta f -\Omega \partial_\phi \delta f = -qA(F) \left(E_{\rm x} \cos\phi + E_{\rm y} \sin\phi\right) \ ,
 \end{equation}
 with $A(F)=\left({kv_\perp \over \omega} \partial_{p_\parallel} F + (1-{k v_\parallel \over \omega})\partial_{p_\perp} F\right)$. The solution of the differential Eq. in $\phi$ is:
 
 \begin{equation}
     \delta f= \delta f^+ + \delta f^-= -{i\over 2}q A(F) \left({(E_{\rm x}-iE_{\rm y})\over \omega-kv_\parallel+\Omega}\exp(i\phi) +{(E_{\rm x}+iE_{\rm y}) \over \omega-kv_\parallel -\Omega}\exp(-i\phi) \right)\ .
 \end{equation}
 Only the perpendicular perturbed current contributes to Lorentz force in the fluid momentum Eq. hence we have \footnote{the terms proportional to e.g. $\exp(i2\phi)$ vanish while we integrate them over $\phi$ as $d^3\vec{p}=d\phi p_\perp dp_\perp dp_\parallel$.} :
 \begin{eqnarray}\label{Eq:DELJ}
     \delta \vec{J}&=& \int d^3\vec{p} \delta f \vec{v}_\perp =-i{q^2 \over 4} \int  d^3\vec{p} v_\perp  \times \\
     &&\left({(E_{\rm x}-iE_{\rm y})\over \omega-kv_\parallel+\Omega}A(F)(\vec{x}+i\vec{y}) +{(E_{\rm x}+iE_{\rm y}) \over \omega-kv_\parallel -\Omega}A(F)(\vec{x}-i\vec{y}) \right)\ , \nonumber
 \end{eqnarray}
 where we have used $\vec{v}_\perp=v_\perp(\cos\phi \vec{x}+\sin\phi \vec{y})$.\\
 Let us express the term $(E_{\rm x}-iE_{\rm y}) (\vec{x}+i\vec{y})$. We have 
 \[
(E_{\rm x}-iE_{\rm y}) (\vec{x}+i\vec{y}) = E_{\rm x} \vec{x}+E_{\rm y} \vec{y} -i E_{\rm y} \vec{x} + i E_{\rm x} \vec{y}
 \]
 If we set $E_{\rm y}=i E_{\rm x}$ (corresponding to $\vec{E}=E(\vec{x}+i\vec{y})$) this terms reads
 \[
 (E_{\rm x}-iE_{\rm y}) (\vec{x}+i\vec{y}) = 2 E_{\rm x} (\vec{x}+i\vec{y}) = 2 \vec{E}^+ 
 \]
 Imposing the same condition leads the second term 
 $(E_{\rm x}+iE_{\rm y}) (\vec{x}-i\vec{y})$ to vanish. Similarly we have for $E_{\rm y}=-i E_{\rm x}$
 \[
 (E_{\rm x}+iE_{\rm y}) (\vec{x}-i\vec{y}) = 2 E_{\rm x} (\vec{x}-i\vec{y}) = 2 \vec{E}^- 
 \]
 Hence we can decompose the perturbed current in Eq. \ref{Eq:DELJ} in the sum of two terms  
 \begin{eqnarray}
 \delta \vec{J}^+ &=& -i{q^2 \over 2} \vec{E}^+\int  d^3\vec{p} {v_\perp A(F) \over \omega-kv_\parallel+\Omega}  \ , \\
 \delta \vec{J}^- &=& -i{q^2 \over 2} \vec{E}^-\int  d^3\vec{p} {v_\perp A(F) \over \omega-kv_\parallel-\Omega}  \ .
 \end{eqnarray}
 These two perturbed currents can be inserted in Eq. \ref{Eq:ED2} back to give in a concise form
 \begin{equation}
     \omega^2= k^2 V_{\rm a}^2 \mp \Omega_{\rm c} {n_{\rm CR} \over n_p} \left(\omega-ku_{\rm CR, \parallel}\right) -{2\pi q^2 V_{\rm a}^2 \omega \over c^2} I^{\pm}(\omega, k) \ . 
 \end{equation}
 where $\Omega_{\rm c} = qB_0/mc$ and $I^{\pm}=\int d^3\vec{p} {v_\perp A(F) \over \omega-kv_\parallel \pm \Omega}$.\\
 
 We now calculate the term $A(F)$ is a special frame moving with a speed $u_{\rm CR, \parallel}$. In this frame the particle angular scattering is sufficiently fast to isotropise the distribution. We further write the distribution in this frame as $F'(p')=F(p_\parallel, p_\perp) E/E'$, where $E$ is the total particle energy. The momentum transformation is $p'_\perp=p_\perp$ and $p'_\parallel=\left(p_\parallel - {E \over c^2} u_{\rm CR, \parallel}\right)/\sqrt{1-u_{\rm CR, \parallel}^2/c^2}$. We now have to convert $A(F)$ in this frame leading to:
 \begin{eqnarray}
     A(F) &=& \left({kv_\perp \over \omega} \partial_{p_\parallel} F + (1-{k v_\parallel \over \omega})\partial_{p_\perp} F\right) \nonumber \\
     &=& \left({kv_\perp \over \omega p'}  \left(p_\parallel-{u_{\rm CR, \parallel} E \over c^2}\right)  + \left(1-{k v_\parallel \over \omega}\right) {p_\perp \over p'} \right){dF(p') \over dp'} \nonumber \\
     &=& {p_\perp \over p'} \left(1- {k u_{\rm CR, \parallel} \over \omega}\right){dF(p') \over dp'}
 \end{eqnarray}
 We have assumed $u_{\rm CR, \parallel} \ll c$ and dropped terms of higher order in $u_{\rm CR, \parallel}/c$, so $F(p') \simeq F(p_\parallel, p_\perp)$.\\
 We now turn to the calculation of the terms $I^{\pm}$.
 \begin{equation}\label{Eq:IP}
    I^{\pm}=\left(1- {k u_{\rm CR, \parallel} \over \omega}\right) \int d^3\vec{p} {dF(p') \over dp'} {v_\perp p_\perp \over p'(\omega-kv_\parallel \pm \Omega)} 
 \end{equation}
 We have $d^3\vec{p}=d^3 \vec{p}' E/E'$ \footnote{$F(\vec{p}) d^3 \vec{p}$ and $d^3 \vec{p}/E$ are both Lorentz invariant (see Landau \& Lifshitz book).}. We make a further simplification considering that the drift in momentum $mu_{\rm CR, \parallel}$ is always lower than the minimum momentum of the particle distribution. This allows us to work at the lowest $u_{\rm CR, \parallel}/c$ order in the equations and to short cut the complicated calculation necessary to express $v_\perp$, $v_\parallel$, and $p_\perp$ in the drift frame in Eq. \ref{Eq:IP} and then drop all prime terms in this Eq and $d^3\vec{p}\simeq d^3 \vec{p}'$. $I^{\pm}$ can be split into two terms $I_1^{\pm}$ and $I_2^{\pm}$ to properly treat the pole in Eq. \ref{Eq:IP} reading as $I^{\pm}=I_1^{\pm} + i I_2^{\pm}$ \footnote{we have used the residue theorem for a pole located on the real axis, the integration contour goes over a half circle giving i$\pi$ times the value of the analytical part at the pole position.} with 
 \begin{equation}
     I_1^{\pm}=-2\pi \left(1- {k u_{\rm CR, \parallel} \over \omega} \right) {\cal P} \int dp d\mu p^2 v {dF(p) \over dp} {(1-\mu^2) \over (kv_\parallel-\omega \mp \Omega)} \ ,
 \end{equation}
 \begin{equation}
     I_2^{\pm}=-2\pi \left(1- {k u_{\rm CR, \parallel} \over \omega} \right)  \int dp d\mu p^2 v {dF(p) \over dp} (1-\mu^2) \pi\delta(kv_\parallel -\omega \mp \Omega) 
 \end{equation}
 where $v_\perp = v \sqrt{1-\mu^2}$. ${\cal P}$ is the principal part of the integral, quite generically we have 
 \begin{equation}
     {\cal P} \int_{-1}^1 d\mu {g(\mu) \over \mu \mp \mu_r} = \lim_{\epsilon \rightarrow 0^+} \left( \int_{-1}^{\pm \mu_r -\epsilon} d\mu {g(\mu) \over \mu \mp \mu_r} + \int_{\pm \mu_r + \epsilon}^1  d\mu {g(\mu) \over \mu \mp \mu_r} \right)
 \end{equation}
 where $\mu_r =(\Omega \pm \omega)/kv$. The term $I_1$ can be written as
 \begin{equation}
     I_1^{\pm} = {2\pi \over k} \left(1- {k u_{\rm CR, \parallel} \over \omega} \right) \int dp p^2  {dF(p) \over dp}  \left(\pm 2\mu_r + (\mu_r^2-1) {\cal P} \int d\mu {1 \over \mu \mp \mu_r} \right)
 \end{equation}

 We have 
 \[
    {\cal P} \int d\mu {1 \over \mu \mp \mu_r} = \log\left({1 \mp \mu_r \over 1 \pm \mu_r}\right)
 \]
 and 
 \begin{equation}
     I_1^{\pm}(k,\omega) = {2\pi \over k} \left(1- {k u_{\rm CR, \parallel} \over \omega} \right) \int dp p^2  {dF(p) \over dp}  \left(\pm 2\mu_r + (\mu_r^2-1) \log\left({1 \mp \mu_r \over 1 \pm \mu_r}\right) \right),
 \end{equation} 
 or 
 \[
 I_1^{\pm}(k,\omega) = \pm {2\pi \over k} \left(1- {k u_{\rm CR, \parallel} \over \omega} \right) \int dp p^2  {dF(p) \over dp}  \left(2\mu_r + (1-\mu_r^2) \log\left|{1 + \mu_r \over 1 - \mu_r}\right| \right) .
 \]
 
 The second term reads 
 \begin{equation}
     I_2^{\pm}(k,\omega) =-{2\pi^2 \over k} \left(1- {k u_{\rm CR, \parallel} \over \omega} \right) \int dp p^2  {dF(p) \over dp} (1-\mu_r^2) \ .
 \end{equation}
The resonant pitch-angle cosine can be expressed in terms of the particle momentum $\mu_r=m (\Omega_c \pm \gamma \omega)/kp$, where $\Omega_c= qB/mc$ is the cyclotron frequency. The perturbations under consideration have a low frequency and we can approximate the resonant pitch-angle cosine as $\mu_r \simeq m\Omega_c/kp$ \footnote{The condition is not valid as $\mu_r \sim 0$. In principle we could keep the general form of the resonant pitch-angle cosine unchanged but then the derivation of the growth rate is more complicated. In order to keep these calculations as simple as possible we then use the approximate form of $\mu_r$.}. We further note $p_r = m|\Omega_c|/k$. \\
Notice that as $|\mu_r| < 1$, $I_2$ involves an integration from $p_r$ to a maximum momentum $p_{\rm max} \gg p_r$. Finally, notice also that the CR distribution $F(p)$ and density $n_{\rm CR}$ are linked together by $n_{\rm CR} \int 4\pi p^2 F(p) dp$.

\paragraph{Dispersion relation and growth rate derivation}
The dispersion relation is 


\begin{equation}\label{Eq:DISI}
      \omega^2= k^2 V_{\rm a}^2 \mp \Omega_{\rm c} {n_{\rm CR} \over n_p} (\omega-ku_{\rm CR, \parallel}) -{2\pi q^2 V_{\rm a}^2 \omega \over c^2} \left(I_1^{\pm}(\omega, k) + i I_2^{\pm}(\omega,k) \right)\ . 
 \end{equation}
 
 If we consider two CR species with the same mass but opposite charges, e.g. electrons and positrons,  the resonant pitch-angle cosine is $\mu_{\rm r,e}=-\mu_{\rm r,e^+}$. If the two species have the same drift speed and densities. Then, $I_{1,e}^\pm = -I_{1,e^+}^{\pm}$ and $I_{2,e}^\pm = I_{2,e^+}^\pm$. This means that if CRs have a vanishing current the $I_1$ (non-resonant) contribution as well as the second term in RHS of Eq. \ref{Eq:DISI} (as $\Omega_{c,e}=-\Omega_{c,e^+}$) to the dispersion relation vanish and only the $I_2$ (resonant) contribution remains. It is important to remind that this contribution comes from an implicit assumption of a population of cold charge in the frame where CR distribution is isotropic which balances their charge excess, for instance in the case CRs are purely protons this term is a contribution of cold electrons. If we have a population of equally charged CRs (electron-positron) then this term comes a contribution of a population of opposite cold components. These if they have the same number do not carry any current and as explained above have a null contribution to the relation dispersion. Amato \& Blasi \cite{ref:ama} show that the derivation in the CR isotropic frame (with a balancing cold electron population) and in the plasma background rest frame (with an excess of thermal electrons) lead to the same dispersion relation in the low frequency perturbation limit.\\
 
 The dispersion relation can be finally cast into the form :
 \begin{equation}\label{Eq:DISPCR}
  \boxed{ \omega^2= k^2 V_{\rm a}^2 \mp \Omega_{\rm c} {n_{\rm CR} \over n_p} (\omega-ku_{\rm CR, \parallel}) \left( 1-T_1(k) \pm i T_2(k) \right) } \ , 
 \end{equation}
 or 
 \begin{equation}
  \omega^2= k^2 V_{\rm a}^2 \pm \Omega_{\rm c} {n_{\rm CR} \over n_p} (\omega-ku_{\rm CR, \parallel}) \left( T_1(k) -1 \mp i T_2(k) \right)  \ , \nonumber 
 \end{equation}
 with
 \begin{equation}\label{Eq:T1}
     T_1 (k)= \int dp {4\pi p^2 F(p) \over n_{\rm CR}} {p_r \over 2 p} \log\left|{p+p_r \over p-p_r}\right| \ ,
 \end{equation}
 and 
 \begin{equation}\label{Eq:T2}
     T_2 (k)=\pi \int_{p_r}^{p_{\rm max}} dp {4\pi p^2 F(p) \over n_{\rm CR}} {p_r \over 2 p} \ .
 \end{equation}
$\pm (T_1(k) -1) -i T_2(k)$ is the function $\zeta_{lr}$ in \cite{ref:zwe}.\\
In order to derive these expressions we have considered that the distribution function $F(p) \rightarrow 0$ as $p \gg p_{\rm max}$ which allows us to drop the supplementary term while proceeding the integration by part over p. 

\paragraph{The resonant branch:}
In case of a vanishing current we set $T_1 = 1$ and the dispersion relation is
\begin{equation}\label{Eq:DISR}
  \omega^2= k^2 V_{\rm a}^2 -i \Omega_{\rm c} {n_{\rm CR} \over n_p} (\omega-ku_{\rm CR, \parallel}) T_2(k)   \ . 
 \end{equation}
The growth rate ($\omega_I$) can be obtained from Eq. \ref{Eq:DISR} by setting $\omega = \omega_R + i \omega_I$. 

\begin{equation}\label{Eq:DIS1}
\omega_R^2 - \omega_I^2=k^2V_a^2 + \Omega_{\rm c} {n_{\rm CR} \over n_p} \bar{\omega}_I T_2 ,
\end{equation}
\begin{equation}\label{Eq:DIS2}
\omega_R \omega_I = - \Omega_{\rm c} {n_{\rm CR} \over 2 n_p} (\omega_R-ku_{\rm CR, \parallel}) T_2  \ .
\end{equation}
If the ratio ${n_{\rm CR} \over n_p} \ll 1$, the solutions are in the test-particle limit. This condition is almost always verified in astrophysical sources \footnote{Some exceptions to this case may be found for instance in hot plasmas close to black holes where non-thermal components dominate the thermal component in density ratios. One may expect strong non-linear feed back from the non-thermal plasma.}. In the test-particle limit the real part is $\bar{\omega}_R \sim k V_{\rm A} \ll k u_{\rm CR, \parallel}$. Alleviating this guess leads to solve an Eq of the 4th degree in $\bar{\omega}_I$. In order to keep the solutions as simple as possible we assume $\bar{\omega}_R = kV_a$ (for the resonant branch only). Using Eq. \ref{Eq:DIS2} (with $\bar{\omega}_R = kV_a$) we find the growth rate (positive $\omega_I$) as 
 \begin{equation}\label{Eq:GRE}
     \Gamma=\bar{\omega}_I \simeq \Omega_c {n_{\rm CR} \over 2 n_p} \left({u_{\rm CR, \parallel} \over V_a}-1\right) T_2(k) \ .
 \end{equation}
 No sign for polarisation appear here, this means that the two circularly-polarised modes (left and right) have the same growth rate. 

\paragraph{The non-resonant branch:} We now set $T_2=0$ in Eq. \ref{Eq:DISPCR} \footnote{If it is always possible to have a vanishing non-thermal current, assuming a vanishing resonant contribution is artificial because as soon as the CR drift speed is larger than the Alfv\'en speed the resonant branch is automatically destabilised. Hence, this discussion is purely pedagogical.} 
The dispersion relation can then be cast into the form:
\begin{eqnarray}
    \left(\omega\pm \Omega_c {n_{\rm CR} \over 2n_p}(1-T_1) \right)^2 &=& \left(k V_a \pm \Omega_c {u_{\rm CR, \parallel} \over V_a} {n_{\rm CR} \over 2 n_p}(1-T_1)\right)^2  \nonumber \\
    &&+ \Omega_c^2 {n^2_{\rm CR} \over 4 n_p^2}(1-T_1)^2 \left(1- {u_{\rm CR, \parallel}^2 \over V_a^2}\right).
\end{eqnarray}
The growth rate is maximal at (using $\omega=\omega_R + i \omega_I$)
\begin{equation}\label{Eq:GNR}
    \Gamma=\omega_I \simeq \Omega_c {u_{\rm CR, \parallel} \over V_a}  {n_{\rm CR} \over 2n_p}(1-T_1) \ .
\end{equation}
 where we have assumed that $u_{\rm CR, \parallel} \gg V_a$. Here a sign appears in the relation dispersion, hence one polarisation is oscillatory (-, $\omega_I < 0$) and the other one growing (+, $\omega_I > 0$). The maximum growth is attained at 
 \begin{equation}
     k= \mp \Omega_c {u_{\rm CR, \parallel} \over V_a^2} {n_{\rm CR} \over 2 n_p}(1-T_1) \ .
 \end{equation}
 Otherwise the growth rate varies as function of k like 
 \begin{equation}\label{Eq:GNRK}
     \Gamma(k) \simeq \sqrt{\Omega_c k u_{\rm CR, \parallel} {n_{\rm CR} \over n_p} (1-T_1) -k^2 V_a^2} \ .
 \end{equation}
 
 \paragraph{General discussion}
 If we have $u_{\rm CR, \parallel} \gg V_a$, Eqs. \ref{Eq:GRE} and \ref{Eq:GNR} are the same for $T_2(k)+T_1(k)=1$. To evaluate at which wave number this does occur one has to fix a particle distribution function. If for simplicity we consider a mono-energetic distribution with $F(p)=(4\pi p_0^2)^{-1} n_{\rm CR} \delta(p-p_0)$, inserting it in Eqs. \ref{Eq:T1} and \ref{Eq:T2} we find $T_1={x \over 2} \ln|(1+x)/(1-x)|$ and $T_2={\pi \over 2} x$  where $x=p_r/p_0$. The above condition is satisfied for $p_0 \sim 2 p_r$. A similar analysis is done in \cite{ref:ama, ref:evo} in the case of a power-law distribution. One can retain that the two growth rates are of the same order for $p \sim p_r$.\\

 As can be seen from Eq. \ref{Eq:GNRK} the non-resonant branch 
 develops between $k_1$ and $k_2$, both wave numbers verify respectively $T_1(k_1)=1$ and $\Omega_c k_2 u_{\rm CR, \parallel} {n_{\rm CR} \over n_p} (1-T_1(k_2)) = k_2^2 V_a^2$. Again for a monoenergetic distribution $k_1$ is obtained for $x \sim 0.8$ or $k_1 r_{g0} \sim 1.25$, where $r_{\rm g0}=p_0/m|\Omega_c|$. The second limit $k_2$ depends on the CR density explicitly. In the limit $k_2 r_{g0} \gg 1$ or $x \ll 1$ we have for a monoenergetic distribution $k_2 \simeq {u_{\rm CR, \parallel} \over 2 V_a} {n_{\rm CR} \over n_p} {\Omega_c \over V_a}$.\\
 
 The above derivation uses a kinetic approach to derive the two growth rates using one formalism. This explains why we used this formalism above. However, in case of cold population of particles the non-resonant growth rate can be derive also adopting a MHD or fluid approach \cite{ref:bell04}. 
 
\subsection{The cold limit : MHD derivation of the growth rate}
This approach has been adopted in the 2004 seminal paper by A.R. Bell \cite{ref:bell04} for a CR current parallel to the background magnetic field and then generalised to any current orientation in \cite{ref:bell05}. The procedure is the same the one leading to Eq. \ref{EQ:CRMHDs}. In the incompressible case and neglecting pressure gradient terms we have the perturbed system:
\begin{eqnarray}
\partial_t \vec{u}&=& -{1 \over 4\pi \rho} \vec{B} \wedge \left(\vec{\nabla} \wedge \delta \vec{B}\right)-{\vec{J} \over \rho c} \wedge \delta \vec{B} - {\delta \vec{J} \over \rho c} \wedge \vec{B}+ {n_{\rm CR} e \vec{u} \over \rho c} \wedge \vec{B} \nonumber \\
\partial_t \delta \vec{B}&=& \vec{\nabla} \wedge \left(\vec{u} \wedge \vec{B}\right) \ .
\end{eqnarray}
If we take $\vec{B}=B \vec{e}_z$, $\vec{J}= J \vec{e}_z$ we find by taking the time derivative of induction Eq. and considering only derivatives along z the same Eq. as Bell's Eq. 3 namely
\begin{eqnarray}
    \partial_t^2 \delta \vec{B}&=& V_{a}^2 \partial_z^2 \delta \vec{B} -{B \over \rho c} \vec{J} \wedge \partial_z \delta \vec{B} -{B \over \rho c} \partial_z \delta \vec{J} \wedge \vec{B} \nonumber \\
    && + {J \over \rho c u_{\rm CR}} \partial_t \delta \vec{B} \wedge \vec {B} \ .
\end{eqnarray}
If we omit the perturbed current term (putting $\sigma \equiv 0$ in \cite{ref:bell04}) we find a dispersion relation like 
\begin{equation}
      \omega^2= k^2 V_{\rm a}^2 \mp \Omega_{\rm c} {n_{\rm CR} \over n_p} (\omega-ku_{\rm CR})\ . \nonumber
 \end{equation}
Bell \cite{ref:bell05} proposes a general calculation with a current not necessarily oriented along the background magnetic field. His analysis includes now density and pressure perturbations because magnetosonic like perturbations can be destabilised but still omitting perturbed current terms. The calculation is performed in the appendix A of \cite{ref:bell05} and is not reproduced here. In the case of non-colinear unperturbed magnetic field and current the fluid has also a non perturbed velocity increasing with time as $\vec{u}_0= - {1 \over \rho} \vec{J} \wedge \vec{B} t$.
The growth rate is given by an Eq. of the sixth order
\begin{eqnarray}
\left(\Gamma^2 \cos^2(\alpha_k)k^2 v_a^2\right)\left(\Gamma^4 +\Gamma^2k^2(v_a^2+ c_s^2)+\cos^2(\alpha_k) v_a^2 c_s^2 \right) = \Gamma_0^4 \times &&\nonumber \\
\left(\Gamma^2 + \cos^2(\alpha_j)k^2c_s^2+k^2v_a^2 (\cos^2(\alpha_k)+\cos^2(\alpha_j)-2\cos(\alpha_j)\cos(\alpha_k)\cos(\alpha_b)\right)&& \nonumber 
\end{eqnarray}
where $\cos(\alpha_k) =\cos(\vec{k},\vec{B})$, $\cos(\alpha_j) =\cos(\vec{k},\vec{J})$, $\cos(\alpha_b) =\cos(\vec{J},\vec{B})$
and $\Gamma_0^4= (\vec{k}.\vec{B})^2 {J^2 \over \rho^2}$, $\Gamma_0$ is the parallel growth rate. When all cosine are set to 1 one can recover the parallel case investigated in \cite{ref:bell04}. The non-resonant streaming instability can be triggered for any orientation of the current $\vec{J}$ with respect to $\vec{B}$ except for $\cos(\alpha_k)=0$, and the growth rate is the strongest for $\cos(\alpha_k)=1$.


\subsection{Thermal effects}\label{S:TEM}
The linear analysis in section \ref{S:LINEM} assumed a cold background plasma. If we want to account for thermal effects the calculation fully involves a kinetic approach using a Vlasov equation for all species (thermal and non-thermal).


Some cumbersome calculations provide a general form of the dispersion relation for parallel perturbations (see KT73).

\begin{equation}
 1-{k^2c^2 \over \omega^2}+ {4\pi \over \omega} \left(i\sigma_{xx} \pm \sigma_{xy}\right) =0 \ .  \nonumber 
\end{equation}
The contribution of the thermal plasma reads as:
\begin{equation}\label{Eq:DEM}
\boxed{ 1-{k^2c^2 \over \omega^2}- \sum_a {\omega_{p,a}^2 \over \omega^2} \left({(\Theta_{a,\parallel}-\Theta_{a,\perp}) \over 2\Theta_{a,\perp}} Z'(\zeta_a^\pm)+ \sqrt{\Theta_{a,\parallel}}(u_{a,d}-{\omega \over k}) Z(\zeta_a^\pm)\right) =0.  }
\end{equation}
The plasma frequency is $\omega_p = \sqrt{4\pi q^2 n/m}$. We have noted $\psi^{\pm}=\psi \pm \Omega_{c,a}$, $\psi_2^{\pm}=\psi \pm 2\Omega_{c,a}$. We introduce the Fried-Conte function 
\begin{equation}\label{Eq:FC}
    Z(\zeta)= {1 \over \sqrt{\pi}} \int du {\exp(-u^2) \over (u-\zeta)}
\end{equation} 
and 
\begin{equation}
    Z_n(\zeta)= {1 \over \sqrt{\pi}} \int du u^n {\exp(-u^2) \over (u-\zeta)} \ .
\end{equation}
and its derivative where $Z'(\zeta)=-2Z_1(\zeta)$ and $\zeta^\pm={\sqrt{\Theta_\parallel}\over k}(\omega-ku_d\pm \Omega_c)$.\\

It then requires to add the contribution of CRs to this equation, this can easily be done using the CR term in Eq. \ref{Eq:DISPCR}. Thermal effects have been analysed in several works : \cite{ref:ach81, ref:rev07, ref:zwe, ref:mar}. All these works considered the case $\Theta_\parallel = \Theta_\perp$ which simplifies Eq. \ref{Eq:DEM} notably. The above expression assumes that the thermal plasma is drifting with a speed $u_{a,d}$ for each species. In the case of CR-driven streaming instability we can either choose the frame in which the thermal plasma is at rest \cite{ref:zwe} or perform the calculation in the frame in which CR have an isotropic distribution \cite{ref:ama}, hence imposing a drift $-u_{\rm CR}$ to the background plasma. We will adopt the former in the following.


The species a are electron/proton background plasma and a component of Cosmic Rays (unless otherwise specified these are protons). For the CR susceptibility we have (see the derivation of the CR part of Eq. \ref{Eq:DISPCR})
\begin{equation}
    X_{\rm CR} = \mp {c^2 \over V_a^2} \left({n_{\rm CR} \over n_i} \Omega_{cp} {(\omega-ku_{\rm CR}) \over \omega^2} \left(T_1\mp iT_2\right)\right) \ .
\end{equation}
The background contribution is calculated assuming an isotropic temperature so $\Theta_\parallel = \Theta_\perp$. In \ref{Eq:DEM} we assumed that no CRs are present. But if CRs are present it is modified to account for both electroneutrality $n_e = n_p + n_{\rm CR}$ (with $n_{\rm CR} \ll n_p$) and background electrons compensating the CR current, hence 
$\vec{J}_{\rm CR}/e = n_{\rm CR} \vec{u}_{\rm CR} = n_e \vec{u}_e$, the CR current imposes a compensating drift to background electrons.
Hence we have $\vec{u}_e \simeq {n_{\rm CR} \over n_p} \vec{u}_{\rm CR}$. It is also possible to define a population of cold electrons drifting with CRs which neutralise the CR charge \cite{ref:ama, ref:zwe}.\\

In order to derive the background contribution we have to account for thermal effects. The cold regime is obtained for $|\zeta^{\pm}| \gg 1$ ($v_{\rm th \parallel} \rightarrow 0)$ and the hot regime $|\zeta^{\pm}| \ll 1$ \cite{ref:mar}. Zweibel \& Everett \cite{ref:zwe} considered the intermediate warm regime with $|\zeta^{\pm}| \gtrsim 1$.\\

\subsubsection{The cold regime}
In the cold case we use $Z(\zeta)^\pm \simeq -{1 \over \zeta^\pm}= {-kv_t \over (\omega -k u \pm \Omega_c)}$ in Eq. \ref{Eq:DEM}. This leads to the dispersion relation in the background thermal plasma using 1) $n_e = n_p + n_{\rm CR}$, 2) $u_e \simeq {n_{\rm CR} \over n_p} u_{\rm CR}$ (and $u_p=0$), we find 
\begin{equation}
    X_{\rm bg} \simeq {c^2 \over v_a^2} \left(1 \pm {\Omega_{cp} \over \omega^2} {n_{\rm CR} \over n_p} (\omega -k u_{\rm CR}) \right) .
\end{equation}
Above we have neglected terms scaling as $m_e/m_p$ and $n_{\rm CR}/n_p$ in front of 1.
Inserting both susceptibilities into Eq. \ref{Eq:SUS} we almost recover Eq. \ref{Eq:DISPCR} \footnote{almost because in the derivation of Eq. \ref{Eq:DISPCR} we dropped the displacement current in the Amp\`ere Eq. hence the term  $V_a^2/c^2$.}
\begin{equation}
 \omega^2 (1+{V_a^2 \over c^2}) \pm \Omega_{cp} {n_{\rm CR} \over n_p}(\omega -k u_{\rm CR}) \left( 1-T_1(k) \pm i T_2(k) \right) - k^2 V_a^2 = 0 \ .
\end{equation}
In the non-relativistic regime the small correction in $V_a^2/c^2$ is neglected. This expression also assumes that the CR density is $n_{\rm CR} \ll n_p$, so is valid in the test particle limit only. Considering only right handed polarised modes (+ sign) the expression corresponds to the case investigated by \cite{ref:bell04} with $T_2=0$.

\subsubsection{The warm background proton regime}
In the warm regime we can perform a Taylor expansion of $u-\zeta^\pm$ in the expression of $Z(\zeta^\pm)$ in Eq.\ref{Eq:FC}. We have to account for the pole $u=\zeta^\pm$ we finally find
\begin{equation}
    Z(\zeta^\pm) \simeq -{1 \over \zeta^\pm} \left[1+{1 \over 2 \zeta^{\pm^2}} \right]+i\sqrt{\pi} \exp(-\zeta^{\pm^2})  \ .
\end{equation}
As noticed by \cite{ref:zwe} the two terms in this expression describe an effect due to a finite gyroradius (the correction is ${k^2r_{\rm th}^2 \over 2(\omega/\Omega_c\pm 1)^2}$, with $r_{\rm th}=V_{T}/\Omega_c$ and $V_T$ is the thermal speed) of the thermal particles (called gyroviscosity) and cyclotron resonance term represented by the imaginary part. Zweibel \& Everett \cite{ref:zwe} have investigated the case of a non-relativistic solution with warm protons and cold electrons. The background susceptibility now reads
\begin{eqnarray}
    X_{\rm bg} &\simeq& {c^2 \over v_a^2} \left(1 \pm {\Omega_{cp} \over \omega^2} {n_{\rm CR} \over n_i} (\omega -k u_{\rm CR}) \nonumber \mp {(k V_{Tp})^2 \over 2 \omega \omega_{cp}}+i\sqrt{\pi} {\Omega_{cp}^2 \over \omega k V_{Tp}} \exp(-{\Omega_{cp}^2 \over k^2 V_{Tp}^2}) \right) .
\end{eqnarray}
hence the dispersion relation reads (still using $u_p=0$ in Eq. \ref{Eq:DEM})
\begin{eqnarray}
 \omega^2  \pm \Omega_{cp} {n_{\rm CR} \over n_p}(\omega -k u_{\rm CR})\left( 1-T_1(k) \pm i T_2(k) \right) - k^2 V_a^2  && \nonumber \\
 + \omega \left(i\sqrt{\pi} {\Omega_{cp}^2 \over kV_{Tp}} \exp\right(-{\Omega_{cp}^2 \over k^2V_{Tp}^2}\left) \mp {k^2V_{Tp}^2 \over 2 \Omega_{cp}} \right)= 0 \ .
\end{eqnarray}
The last bracket shows the thermal corrections due to warm ions.
Zweibel \& Everett \cite{ref:zwe} derived the maximum growth rate and the wavenumber for the non-resonant branch (upper sign, $T_2 =0$) and get
\begin{eqnarray}
\Gamma(k) \simeq \Omega_{cp} \left({n_{\rm CR} \over n_i}{u_{\rm CR} \over V_{Tp}}\right)^{2/3} \\
k_{\rm max} \simeq {\Omega_{cp} \over V_{Tp}} \left({n_{\rm CR} \over n_p}{u_{\rm CR} \over V_{Tp}}\right)^{1/3} \ .
\end{eqnarray}

\subsubsection{The hot background proton regime}
In the hot case we use $|\zeta|^{\pm} \ll 1$ \cite{ref:mar} and the Fried-Conte functions are (the principal part $Z(0) = 0$ because the integrand is an odd function) 
\begin{equation}
    Z(\zeta) \simeq i\sqrt{\pi} \exp(-\zeta^2)-2 \zeta \left(1-{2 \over 3} \zeta + o(\zeta^2)\right) \ .
\end{equation}
In this regime the thermal protons are said to be de-magnetised, this means that their Larmor radius is larger than the scale of the perturbations, i.e. $kr_{\rm th} \gg 1$. To the lowest order in $\omega/\Omega_c$ we find
\begin{eqnarray}
\mp \Omega_{cp} {n_{\rm CR} \over n_p}(\omega -k u_{\rm CR})\left( T_1(k) \mp i T_2(k) \right) -k^2 V_a^2 && \nonumber \\ 
+ \omega \left(\mp 2 {\Omega_{ci}^3 \over k^2 V_{Tp}^2} + i\sqrt{\pi} {\Omega_{ci}^2 \over kV_{Tp}}\right)  \nonumber \\
\pm \Omega_{cp} \left(\omega - ku_{\rm CR} {n_{\rm CR} \over n_p}\right) = 0 \ ,
\end{eqnarray}
where we have assumed $\exp(-\zeta^2) \simeq 1$. Cold electrons contribute to the third row while hot protons to the second. This result is at $o(({n_{\rm CR} \over n_p})^2)$ and we have neglected 
a term in the electron susceptibility scaling as ${m_e \over m_p}$.\\

In the case of the non-resonant mode only, for CRs with a Larmor radius $r_{\rm L} \gg 1/k$, at the maximum growth rate Marret et al \cite{ref:mar} find a growth rate 
\begin{equation}
    \Gamma(k) \simeq {n_{\rm CR} \over n_p} {u_{\rm CR} \over V_{Tp}} \Omega_{cp} \ .
 \end{equation}
\subsection{The streaming instability in partially ionised media}
Astrophysical plasma environments usually are only partially ionised, it is hence important to consider the effect of neutrals over the perturbation propagation \footnote{Recently \cite{ref:squ} also account for the effect of dust particles over CR confinement.}. Notice first that the dispersion relation of MHD waves in a partially ionised medium has been investigated in a series of work see e.g. \cite{ref:sol1, ref:sol2, ref:xu16}. \cite{ref:kul} includes the effect of CRs.\\
These effects can be included in the fluid formalism by adding a friction term in the (ionised) momentum equation, namely rewriting Eq.\ref{EQ:CRMHDs} as
\begin{eqnarray}\label{EQ:CRMHDsn}
\partial_t \vec{u}_i + \vec{u}_i.\vec{\nabla} \vec{u}_i &=& {1\over 4\pi \rho_i} (\vec{\nabla}\wedge \vec{B}) \wedge \vec{B}-{1 \over \rho_i c} \vec{J}_{\rm CR} \wedge \vec{B} - {\vec{\nabla} P_i \over \rho_i} -{e n_{\rm CR} \over \rho_i} \vec{E}  \nonumber \\
&&-\nu_{in} \left(\vec{u}_i - \vec{u}_n \right) \ , \nonumber \\
\partial_t \vec{u}_n + \vec{u}_n.\vec{\nabla} \vec{u}_n &=& - {\vec{\nabla} P_n \over \rho_n} -{\rho_i \over \rho_n} \nu_{in} \left(\vec{u}_i - \vec{u}_n \right) \ , \nonumber \\
{1 \over c} \partial_t \vec{B} &=& -\vec{\nabla} \wedge \vec{E} \ , \nonumber \\
\vec{E} &=& -{\vec{u}_i \over c} \wedge \vec{B} \ .
\end{eqnarray}
Here the index i stems for ions and n for neutrals. We have introduced the ion-neutral collision frequency $\nu_{in}$ which is a function of the local ion temperature, namely (see 
\cite{ref:shu, ref:jea, ref:rec}).
\begin{eqnarray}
\nu_{in} &\simeq& 8.9~10^{-9} {n_n \over 1~\rm{cm^{-3}}} \left({T_i \over 1~\rm{eV}}\right)^{0.4}~\rm{s^{-1}} \ , \rm{for~ 10^2 K < T < 10^6 K} \ , \nonumber \\
\nu_{in} &\simeq& 1.6~10^{-9} {n_n \over 1~\rm{cm^{-3}}}~\rm{s^{-1}} \ , \rm{for~ T < 10^2 K} \ .
\end{eqnarray}
The neutral-ion collision frequency $\nu_{ni}$ can be deduced from
$\rho_i \nu_{in}= \rho_n \nu_{ni}$.\\
We follow on the same procedure as for the ionised case. We express the neutral perturbed velocity from the neutral momentum Eq. as
\begin{equation}
    \vec{u}_n = \left({\chi \nu_{in} \over \chi \nu_{in} -i \omega} \right) \vec{u}_i \ ,
\end{equation}
then still neglecting pressure gradient effects we find instead of Eq. \ref{Eq:ED2}
\begin{equation}\label{Eq:ED2n}
    i\omega \vec{E}= \left({i k^2 V_{\rm a,i}^2 \over \omega} +{\omega \nu_{in} \over i\chi \nu_{in}+ \omega} \right)\vec{E} +{4\pi V_{\rm a, i}^2 \over c^2} \delta \vec{J}_{\rm CR}^{\pm} \mp i{e B_0 n_{\rm CR} \over \rho c} \left(1-{ku_{\rm CR, \parallel} \over \omega}\right)\vec{E} \ ,
\end{equation}
where $\chi = \rho_i/\rho_n$ and the ion Alfv\'en speed is $V_{\rm a,i}= {B_0 \over \sqrt{4\pi \rho_i}}$. The dispersion relation given by Eq. \ref{Eq:DISPCR} is then modified as (see also \cite{ref:rev21})
\begin{equation}\label{Eq:DISPCRn}
  \boxed{ \omega^2 \left( 1 + {i \nu_{in} \over \omega+ i\chi \nu_{in}}\right) = k^2 V_{\rm a, i}^2 \mp \Omega_{\rm c} {n_{\rm CR} \over n_p} (\omega-ku_{\rm CR, \parallel}) \left( 1-T_1(k) \pm i T_2(k) \right) } \ .
 \end{equation}
It is not of prime interest to also include thermal effects in that case as the latter occur at temperatures where the fraction of neutrals drops toward zero, hence the cold approximation is good to investigate the effects of neutrals. The dispersion relation is of order 3, hence usually has three solutions. We introduce $\chi  = {\rho_i \over \rho_n} = {\nu_{ni} \over \nu_{in}}$ as a parameter to evaluate the degree of ionisation in the system. The case $\chi \gg 1$ gives the unmodified dispersion relation. \\

\paragraph{Ion-neutral coupling}
Depending on the ratio between the frequency of a perturbation $\omega$ and the frequencies $\nu_{in}, \nu_{ni}$ in a partially ionised plasma we have a different behaviour of ions with respect to neutrals. Namely, if $\omega \ll \nu_{ni} < \nu_{in}$, neutrals have time to adapt to ion motion and both ions and neutrals are coupled. In this regime perturbations in the ion motions are weakly damped. On contrary if $\omega \gg \nu_{in}$, ions and neutrals are decoupled and the effect of collisions is maximal over ion motions, in this regime the latter are strongly damped. Finally, there is an intermediate regime $\nu_{in} > \omega > \nu_{ni}$, in that case neutral-ion collisions have time to damp the magnetic perturbations but there is no time for a momentum transfer to the neutral fluid.
Alfv\'en waves in this regime can not propagate if $\chi < 1/8$ (\cite{ref:kul, ref:zwe82, ref:bra}).

\paragraph{The non-resonant branch}
Reville et al \cite{ref:rev07} investigate the non-resonant branch $T_2 =0$ and found imaginary roots, two damped and one growing (the non-resonant mode). Hence, neutrals are unable to fully stabilise the non-resonant instability. A simpler solution can be found in the so-called strongly driven case. We introduce here two quantities
the typical CR Larmor radius driving the instability $r_{d}={p_d \over m_p \Omega_c}$ and $\delta = {n_{\rm CR} p_d \over n_p m_p u_{\rm CR}}$. The CR term can then be written as $A_{\rm CR} \simeq k^2 u_{\rm CR}^2 {\delta \over k r_d} (1-T_1)$. The strongly driven case is given by the condition $\delta {u_{\rm CR}^2\over V_{a,i}^2} \gg k r_d$. The non-resonant branch develops for $k r_d \gg 1$. In that case the system reads
\begin{equation}
    \omega^2 \left(\omega + i (\nu_{in} + \nu_{ni})\right) = \pm A_{\rm CR} (\omega + i \nu_{ni}) \ .
\end{equation}

In the low ionisation case, i.e. for $\chi \ll 1$, considering a purely growing mode hence $\omega \simeq i \omega_i$ we find the solution 
\begin{equation}
    \omega_i = {-\nu_{in} \over 2} + {1\over 2} \sqrt{\nu_{in}^2 + 4 A_{\rm CR}} \ .
\end{equation}
The regime $\nu_{in}^2 \gg 4 A_{\rm CR}$ gives 
\begin{equation}
    \omega_i \simeq {A_{\rm CR} \over \nu_{in}} \ ,
\end{equation}
we have $\omega_i \ll \nu_{in}$ \cite{ref:byk05}.

\paragraph{A general analysis}
In a recent work \cite{ref:rev21} a complete re-analysis has been proposed of CR-driven streaming modes in a partially ionised plasma (hence including the resonant branch). It is based on the development of the distribution function in spherical harmonics (see Eq. \ref{Eq:FSH}). In particular the authors retain the anisotropy up to second order terms and investigate the effect of CR pressure anisotropy over the streaming-triggered modes. A pressure anisotropy indeed modifies the wave propagation in the resonant branch. The non-resonant branch is unaffected because linked to the CR current. Reville et al \cite{ref:rev21} argue that CR-driven waves need another approach than solving Eq.\ref{Eq:DISPCRn} in the limit with no CRs. In effect, CR-driven modes in partially ionised medium differ because of two main effects with respect to the propagation of normal Alfv\'en modes: 1) ion motions are forced \cite{ref:dru96} 2) driven waves propagate in an inhomogeneous plasma (the shock precursor for instance). The latter in particular requires to solve for the dispersion relation in $k(\omega)$ rather than in $\omega(k)$, so for real k. In that case as shown by \cite{ref:tag} waves can have an evanescent propagation but do not have region of k where the real part of the frequency does not exist \cite{ref:zwe82}.

\section{Numerical studies}
The physics involved in the late evolution of CR driven instability is highly non-linear and hence requires the use of specific numerical approaches. Unfortunately because of the large dynamical range of scales (spatial / temporal / energetical) involved in the process several complementary techniques are necessary to start to grasp these non-linear multi-scale complex systems. There are basically three main techniques (and one semi-analytical approach) which has been used to describe cosmic rays coupled to the magnetised thermal plasma. We describe these shortly below and highlight some of their "feats of arms".\\

\subsection{Particle-in-cell simulations}
Particle-in-cell (PIC) methods treat the plasma kinetically,
describing both ions and electrons as a collection of particles and solve
the equations of particle motion by iteration under the effect of the
Lorentz force coupled to the Maxwell equations \cite{ref:bir, ref:poh}.
\subsection{Hybrid simulations}
In hybrid methods, electron dynamics are not treated kinetically
but using a fluid model. Ions are treated using PIC techniques \cite{ref:lip}. Non-thermal particles can be included in the PIC solver \cite{ref:gar}.
\subsection{Magneto-hydrodynamic simulations including non-thermal components}
In these approaches, the thermal plasma is treated using MHD
equations, whereas non-thermal particles are treated kinetically using
PIC techniques \cite{ref:bai15, ref:vma}. Alternatively,
non-thermal particles can be treated using a kinetic model like Vlasov \cite{ref:pal} or Vlasov–Fokker–Planck \cite{ref:rev13}.

\subsection{Words of caution}
It is important to keep an important information in mind when we are dealing with numerical simulations: these can be seen as experiments made with a computer, but they remain an experiment. From this type of experiment using a theoretical corpus (see the previous section) we can extract some information to explain its results. As is the case when we want to describe the dynamics of complex systems, as astrophysical/space plasma systems are, the results obtained at the end of a simulation are dependent on the so-called simulation set-up, that is the choice of initial conditions and boundary conditions, spatial and time resolution, geometry and methods. This is why a result from a numerical simulation has to be interpreted within a specific framework. Besides, simulations in particular to what concerns instability studies are not performed under realistic (astrophysical) conditions but often with artificially modified initial conditions in order to be able to track the expected physical effect under a reasonable simulation timescale. The final result has to be properly rescaled to the astrophysical/ space plasma situation under investigation. All the aforementioned issues make the interpretation of a numerical run to be imbued with some amount of caution if we want to generalise it to realistic physical systems.

\subsection{The resonant streaming instability}
\subsubsection{Particle-in-cell studies} 
\begin{figure}
    \centering
    \includegraphics[width=\linewidth]{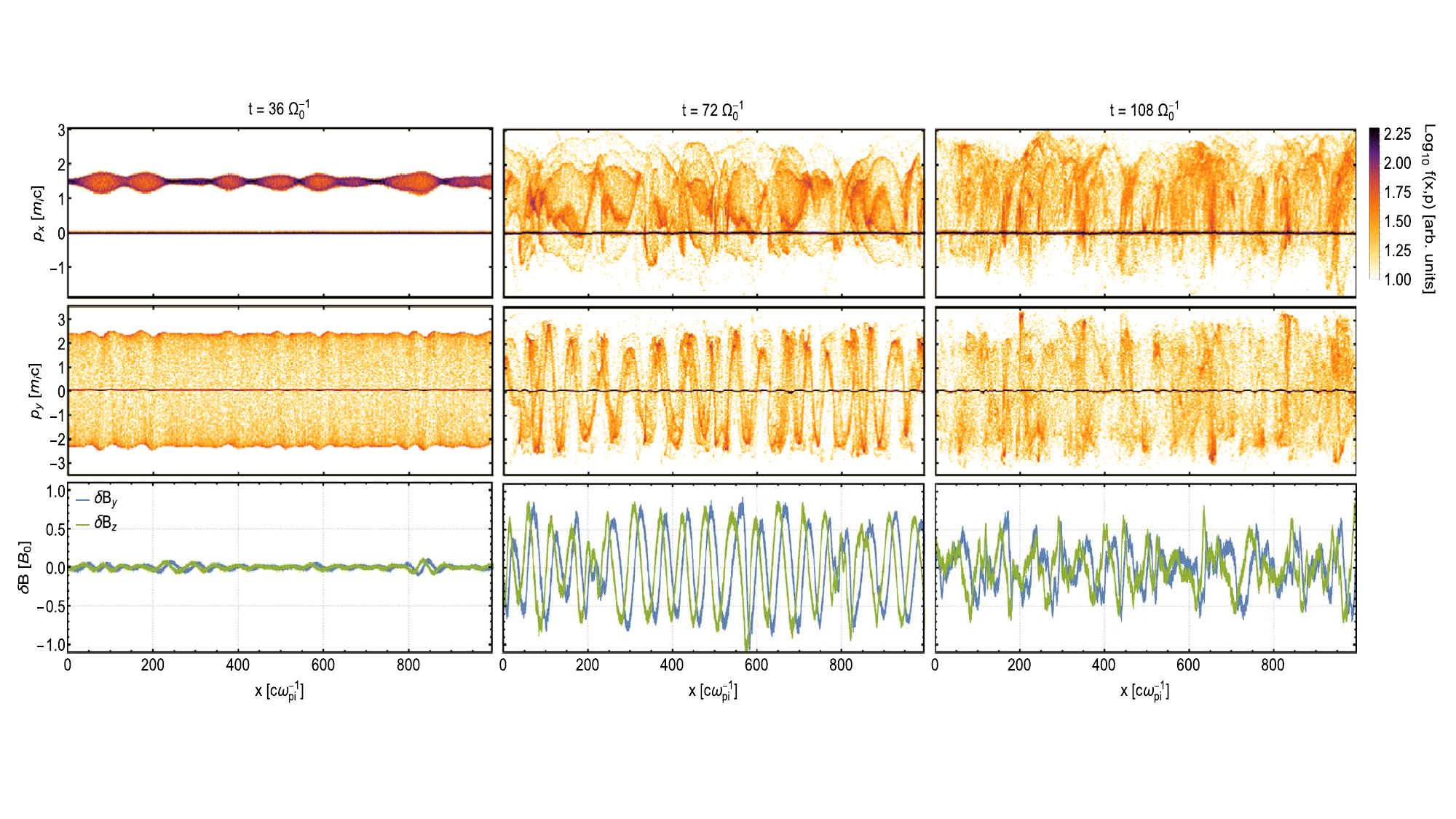}
    \caption{Evolution of the perturbed magnetic energy and CR anisotropy at three times for the RLD case Gy4 (see text for the parameters). By courtesy of \cite{ref:hol}.}
    \label{F:RDFSI}
\end{figure}

Holcomb \& Spitkovsky \cite{ref:hol} propose a study of the behaviour of the resonant instability using PIC simulations.\\ 
{\it simulation set-up:} The simulations are 1D3V, this means that only one spatial dimension is retained whereas the outputs retained the information over the three components of each particle velocity. The authors considered two different type of initial CR (proton) distribution: 1) a monoenergetic gyrotropic ring-like distribution (hereafter RLD), and 2) a power-law distribution (hereafter PLD), isotropic in a frame moving with a drift speed $u_d$ with respect to a background electron-proton plasma. The PLD particle Lorentz factors are sampled in the interval [2, 10]. An advantage of the ring distribution is to isolate the wave–particle resonance in the momentum phase space as both momenta and particle pitch-angle are initially set. The simulation boundary conditions are: periodical with an Alfv\'en speed $V_a =0.1 c$, the initial magnetic field is homogeneous along the x direction, the background ions are cold with a thermal speed $V_{\rm T,i}=0.01 c$. The simulation resolution has an electron skin depth $c \omega_{\rm pe}=10$, the bow length cover 50 to 1000 cells (see their table 1). PIC simulations are conducted with a mass ration ${m_i \over m_e}=100$. The number of particle per cell is equally balance among thermal and Cosmic Rays (so non-thermal to thermal particle density ratio is one), but the mass of CR ion is reduced with respect to the background one so that the effective density ratio $({n_{\rm CR} \over n_i})_{\rm eff}= {m_{\rm CR} \over m_i} \ll 1$. The typical CR drift speed is $u_d = (2-8) V_a$, hence the growth rate is derived in the background plasma rest frame. The simulation maximum time is about 5000 proton gyro periods.\\
{\it Main results - growth rates -} The RLD case is instructive to deconstruct the physics of the instability. We mainly discuss here the results of simulation Gy4 [particular set-up: $u_d = 5 v_a$, $({n_{\rm CR} \over n_i})_{\rm eff}= 2~10^{-2}$, a very long box size with about 100 000 cells]. As can be seen in Fig. \ref{F:RDFSI} the instability growth proceeds like this: at an early timescale (36 $\Omega^{-1}$) a gyroresonant mode is observed in the magnetic pattern, at the end of the linear growth phase a value of $\delta B/B_0 \sim 0.5$ is obtained which brings us beyond the usual QLT, then at later stage pitch-angle scattering induces an isotropisation as seen in $p_x$ and $p_y$ plots. The simulation Gy1 which has $({n_{\rm CR} \over n_i})_{\rm eff}= 10^{-4}$ allows to clearly disentangle the main growing modes, actually the parallel ($k > 0$) right  and anti-parallel ($k < 0$) left with a good agreement with theoretical expectations \cite{ref:she}. 

\begin{figure}
    \centering
    \includegraphics[width=\linewidth]{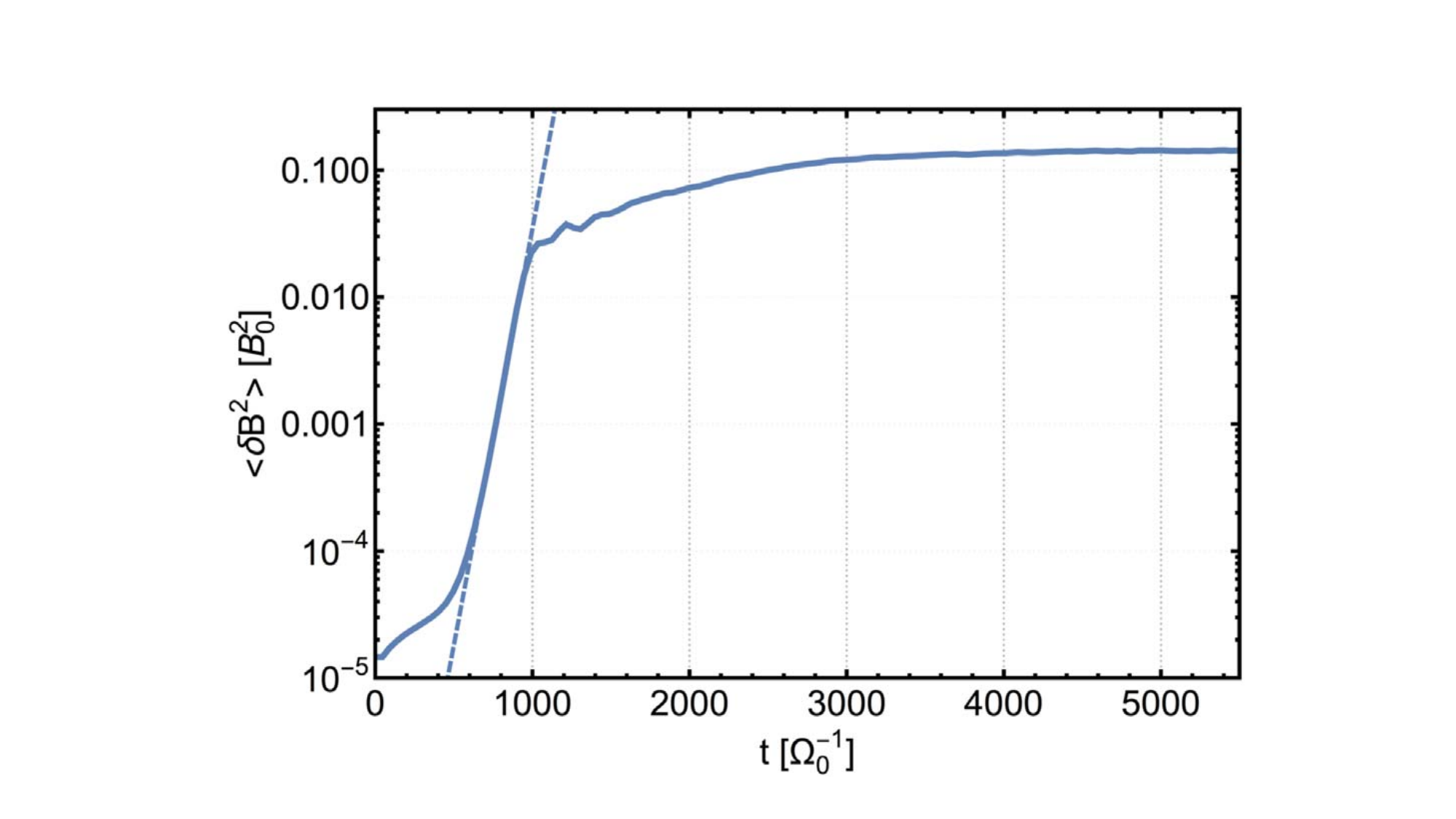}
    \caption{Evolution of the perturbed magnetic energy as function of time for the PLD case Hi3 (see text for the parameters). The first row represents the momentum along the initial magnetic field in the x direction. The second row represents the momentum perpendicular to the initial magnetic field in the y direction. The third row represents the amplitude of the y and z magnetic components. By courtesy of\cite{ref:hol}.}
    \label{F:GRSI}
\end{figure}

In the PLD case, the perturbed magnetic energy for the simulation Hi3 (high anisotropy case) is displayed in Fig. \ref{F:GRSI} [set-up: $u_d = 7.9 V_a$, $({n_{\rm CR} \over n_i})_{\rm eff}= 2~10^{-3}$, a very long box size with more than one million cells]. An estimation of the maximum growth rate can be made using a linear regression (linear curve). It usually leads to an overestimation with respect to the theoretical calculation by a factor up to 2 possibly associated with numerical issues like heating associated with high k noise production.\\

\begin{figure}
    \centering
    \includegraphics[width=1.5\linewidth]{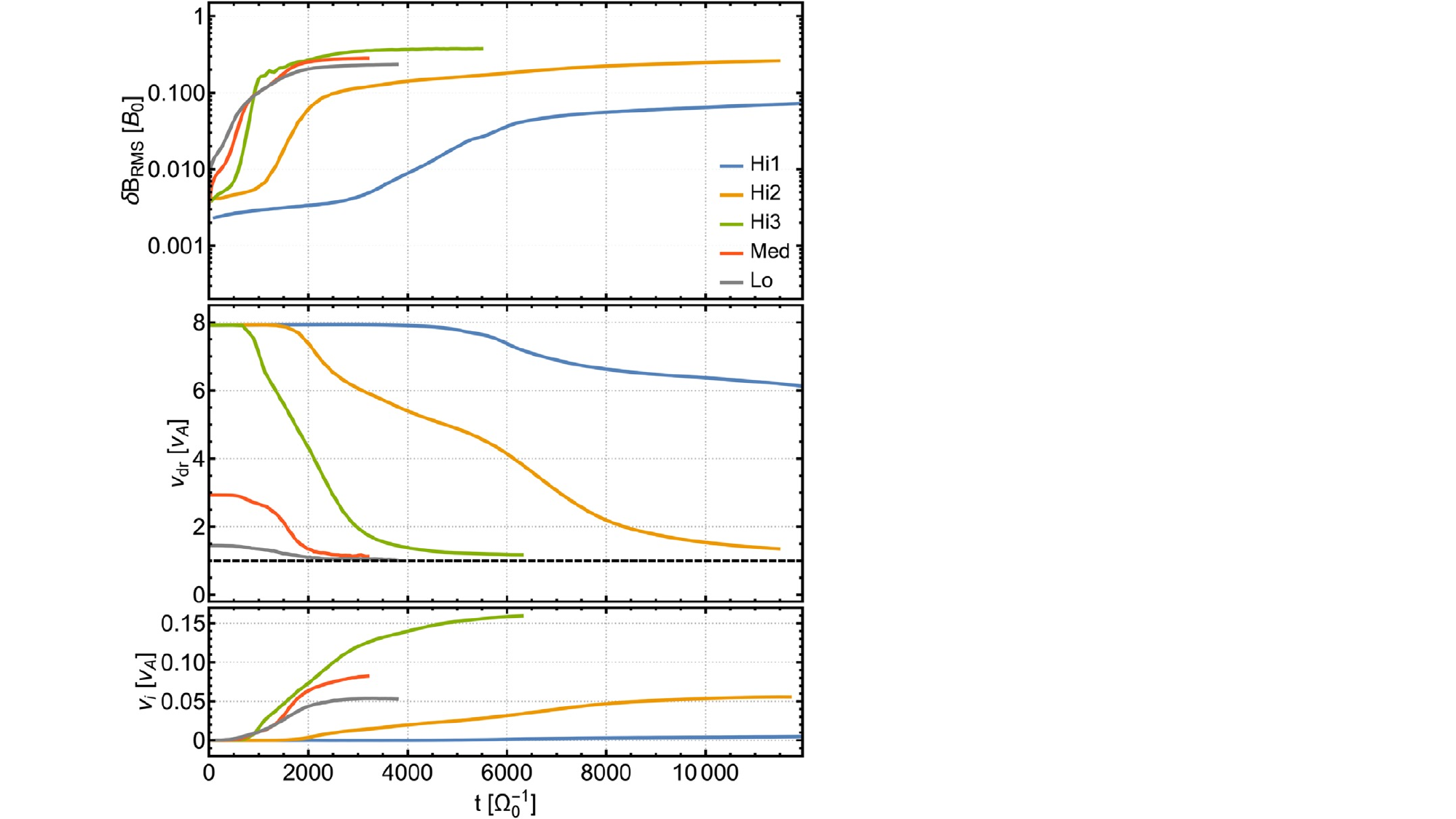}
    \caption{Up: the rms of the perturbed magnetic field as a function of time in the PLD case for different anisotropy strength (from high (Hi) to low (Lo)). Middle: the same but for the CR drift speed. Bottom: the same but for the background ion drift speed. By courtesy of \cite{ref:hol}.}
    \label{F:SIREL}
\end{figure}

{\it Main results - saturation -} 
The saturation process depends on the type of CR initial distribution. Basically, in the RLD the saturation is due to particle trapping inside magnetic mirrors
produced while the amplitude of the perturbed magnetic field raises. In that case, particle distribution starts to isotropise over a timescale corresponding to the inverse of the bouncing frequency $\Omega_t$ given by 
\begin{equation}
    \Omega_t = \sqrt{{\delta B \over B_0} k v_\perp \Omega} \ ,
\end{equation}
where $v_\perp$ is the particle velocity in the direction perpendicular to the background magnetic field. Setting $\Omega_t = \Gamma$ gives a criterion to fix $(\delta B/ B_0)$ at saturation for a given pitch-angle cosine. It appears that depending on the CR density involved in the simulation the isotropisation process can not always be completed because the amplitude of the perturbed field is too low. The isotropisation process is accompanied by a reduction of the drift speed to the local Alfv\'en speed, in fact exactly at a speed $V_a + V_{\rm T,i}$ because background ions acquire some drift in the strongly anisotropic cases. This effect is due to the $\vec{E} \wedge \vec{B}$ imposed over the background ions by the waves. In the PLD case the saturation is more complex as it involves the filling of pitch-angle space by different wave numbers aside the possible mirroring effect which is present only if the amplitude of the perturbed field is high enough. In the highly anisotropic case, the saturation process has two steps (see Fig. \ref{F:SIREL}): a first non-linear growth phase where modes continue to grow at wave number different from the maximum one. In that phase, the isotropisation process around $90^o$ is not completed. In a second phase, enough left-handed modes are generated to allow the CR pitch-angle to pass the $90^o$ barrier and to lead a full particle distribution and then to the relaxation of the CR drift. The CR drift relaxes to $V_a + V_{\rm T,i}$. The drift imposed on ions is possibly at the origin of the wind launching from the galactic disc. 

\subsubsection{Magnetohydrodynamic-kinetic studies}
Bai et al \cite{ref:bai19} conduct an investigation of the resonant streaming instability but using a PIC-MHD approach. \\

{\it simulation set-up:} CR are injected following a $\kappa-$distribution which converges to a power-law distribution for momenta $p \gg p_0$, and flat below $p_0$. We have $p_0 > p_{\rm min}$ the minimum CR momentum. The simulations are conducted in a frame where the CR distribution is close to isotropy letting the background plasma to drift with a speed $\vec{u}_d = - u_d \vec{x}$ where the background initial magnetic field points in the $+\vec{x}$ direction. The drift speed is usually low $2 V_a$, a simulation has $u_d = 8 V_a$. The boundary conditions for gas and particles are periodical. CR particles verify $q/mc = 1$ and the code unit length is $d=V_a/\Omega = 1$. CR to gas density ratios are small and span the range $10^{-5}$ to $10^{-3}$ (see their table 1). The authors adopt a particular approach because of the weak anisotropy of the CR distribution as called the $\delta f$ calculation. In this method a time-dependent weight is affected to each particle which measures its offset with respect to the initial distribution function. This method allows considerable noise reduction. The simulations are 1D3V. The box is very elongated along the $x$-axis to allow CRs to eventually travel it at the speed of light in time larger than the typical growth time of the resonant modes. The maximum timescale is typically about 1 million time ion gyro time. \\

\begin{figure}
    \centering
    \includegraphics[width=1.\linewidth]{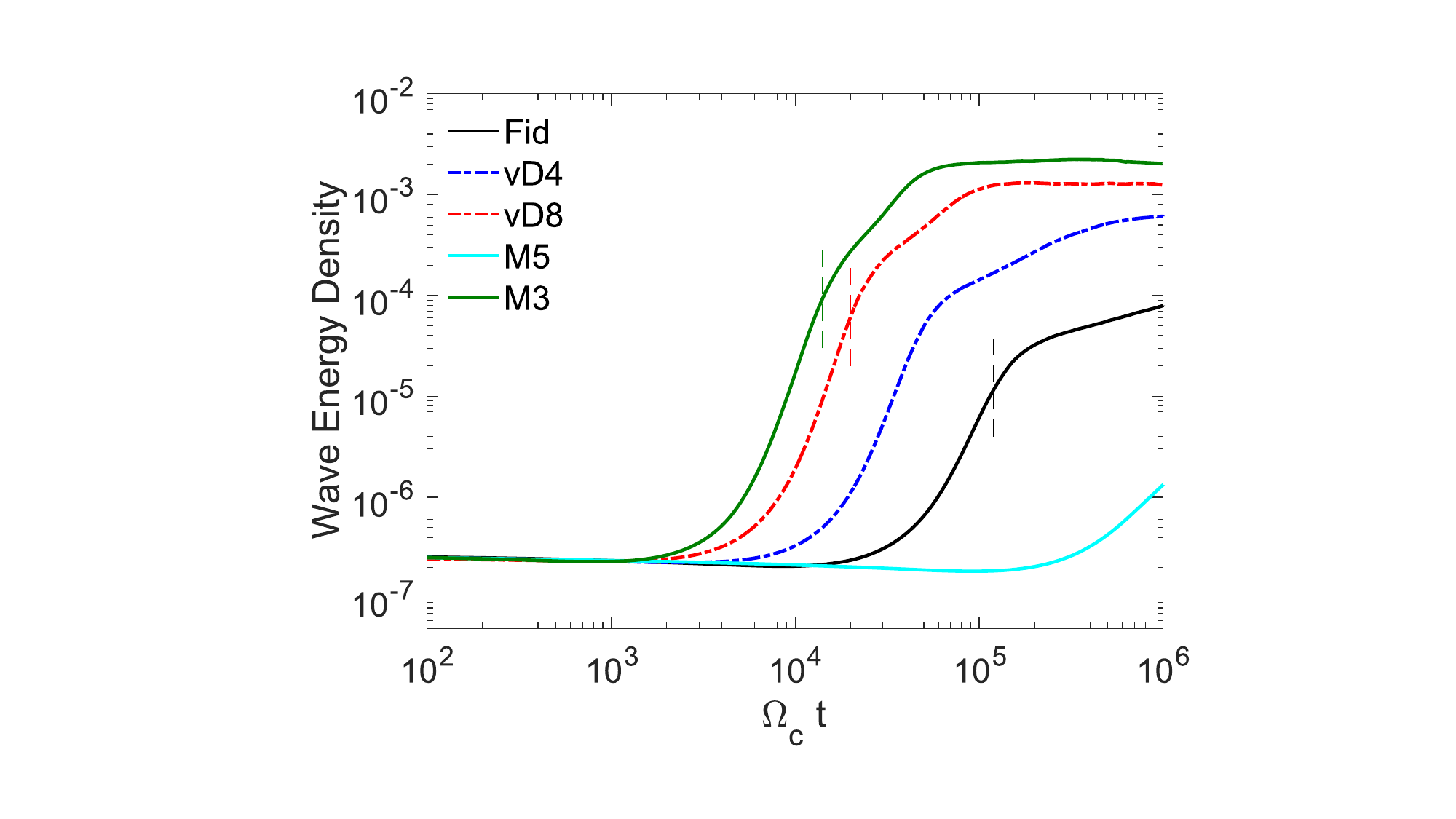}
    \caption{Time evolution of the wave energy density for the different set-ups. The dashed lines mark the end of the linear growth phase. By courtesy of \cite{ref:bai19}.}
    \label{F:DELB}
\end{figure}

{\it Main results - growth rates -}
The perturbed magnetic energy time evolution is plotted in Fig. \ref{F:DELB}. The figure clearly shows the linear phase which ends at the dashed lines. The wave energy still grows after but a moderate rate (see next section) during a non-linear growth phase before reaching saturation. 

\begin{figure}
    \centering
    \includegraphics[width=1.\linewidth]{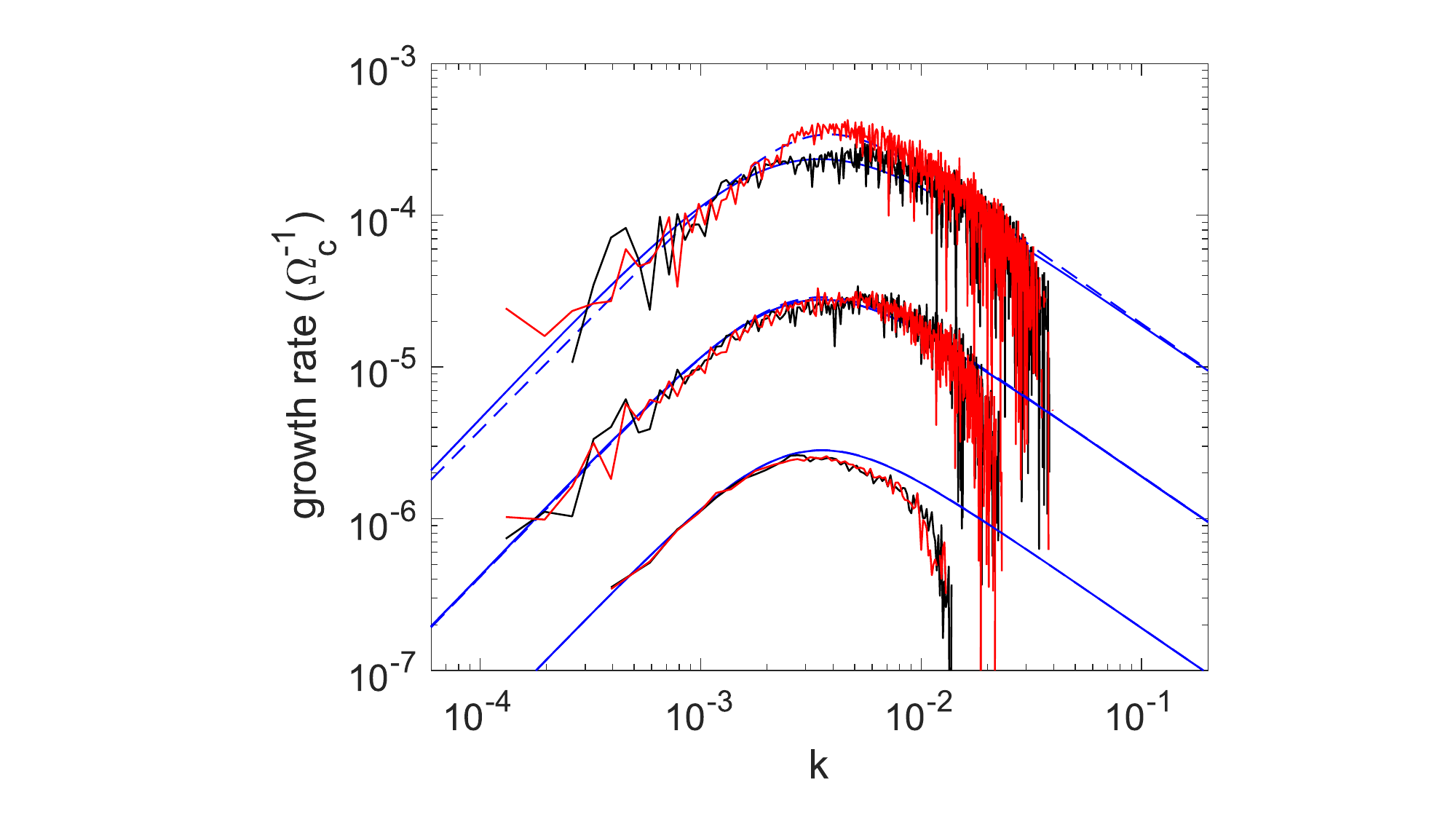}
    \caption{Linear growth rate measured in different set-ups M3 (top), Fid
(middle), and M5 (bottom). The black/red curves mark the left-/right-handed
polarisation waves. Blue solid/dashed lines mark the analytical growth rate
expected from a $\kappa$ distribution for the left-/right-handed modes. By courtesy of  \cite{ref:bai19}.}
    \label{F:GRPOL}
\end{figure}

Because of the presence of particles at all pitch-angle cosine between -1 and 1 and because these particles are protons then forward-traveling CR particles trigger left-polarised, forward-propagating Alfvén waves (FL waves), whereas backward-traveling CR particles excite right-polarised, forward-propagating Alfvén waves (FR waves). This is clearly the case on Fig. \ref{F:GRPOL} which shows the derived growth rate for both FR and FL waves compared with the theoretical prediction in blue for three different set-ups [all have $u_d = 2 V_a$,  M5 : ${n_{\rm CR} \over n_b} =10^{-5}$, Fid: ${n_{\rm CR} \over n_b} =10^{-4}$, M3 : ${n_{\rm CR} \over n_b} =10^{-3}$]. The slight difference appearing at high CR to gas ratio is linked to the (1-$T_1$) in the dispersion relation which modifies the real part of the dispersion relation. As this ratio drops both polarisation are expected to grow with the same rate. 

\begin{figure}
    \centering
    \includegraphics[width=1.\linewidth]{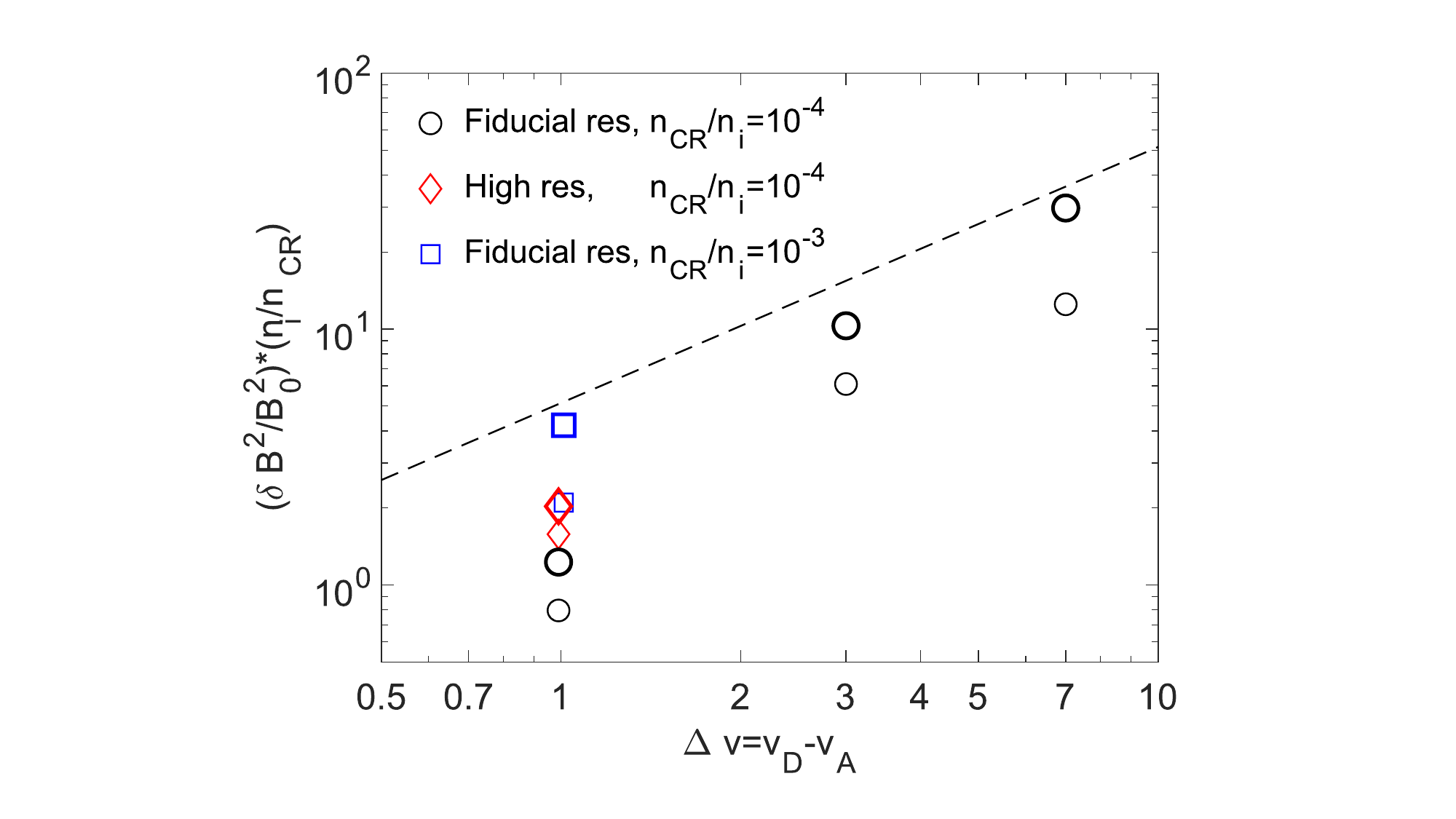}
    \caption{The level of magnetic energy density at the saturation: comparing simulations (symbols) with the theoretical level (dashed line) as a function of the initial drift speed. Thick symbols correct for the numerical dissipation. By courtesy of \cite{ref:bai19}.}
    \label{F:SATPICMHD}
\end{figure}

{\it Main results - saturation -} 
The issue with saturation is strongly linked with the way the particle distribution is isotropised. This is inertly encoded in the resonance condition 
\begin{equation}
    k p \mu = m \Omega_c
\end{equation}
For small $\mu$ the product $k p$ has to be high enough to allow particles of a given momentum p to get scattered efficiently. The process of crossing the $90^o$ barrier is then related to the amount of modes at the right k with right polarisation. The larger k are needed the more difficult the barrier crossing. This effect is also mitigated by the injected distribution which is flat at momenta below a reference one $p_0$ and then drop as a power-law at higher p. All in all, particle isotropisation appears finally easier at $p < p_0$. If pitch-angle cosine scattering is permitted this will fix the saturation process and the amplitude of the saturation level which can be compared to an analytical quasi-linear estimation. Fig \ref{F:SATPICMHD} compares the saturation magnetic field energy density with the theoretical expression. The agreement is reasonably good at all initial drift speeds if numerical dissipation is properly accounted for as can be seen for thick symbols. In the small drift cases, the simulations underestimate the saturation level because of a lack of 90$^o$ crossings. Notice that the theoretical expression considers at total transfer of the initial beam momentum to a class of waves so it is likely an overestimate.  

{\it Environmental effects: ion-neutral damping-} Plotnikov et al \cite{ref:plo}
continue using the same PIC-MHD approach with the $\delta f$ method but in an environment composed of neutrals. In that case one may expect ion-neutral damping to control the saturation of the magnetic field. The simulations however, have been conducted with a one fluid MHD approach. Hence, these simulations are only valid in the regime where ions and neutrals are decoupled (high frequency regime). The fiducial drift speed is 10 $V_a$ in this work otherwise the parameters and set-ups are similar to \cite{ref:bai19} (see their table 2 for completeness). The study tests different ion-neutral collision to maximum growth rate ratios $R=\nu_{\rm in}/\Gamma_{\rm max}$. The results show the following trends: the linear growth rate is strongly suppressed for $R > 1$, converging to zero. The level of saturation of the magnetic field accordingly drops with R, as soon as $R > 10^{-2}$, the saturation starts to be controlled by ion-neutral collisions. Ion-neutral collisions in the decoupled regime damp preferentially high k modes which are necessary to ensure $90^o$ crossing, hence ion-neutral collisions exacerbate the difficulty to isotropise the particle distribution.

\subsection{The non-resonant streaming instability}
\begin{figure}
    \centering
    \includegraphics[width=1.\linewidth]{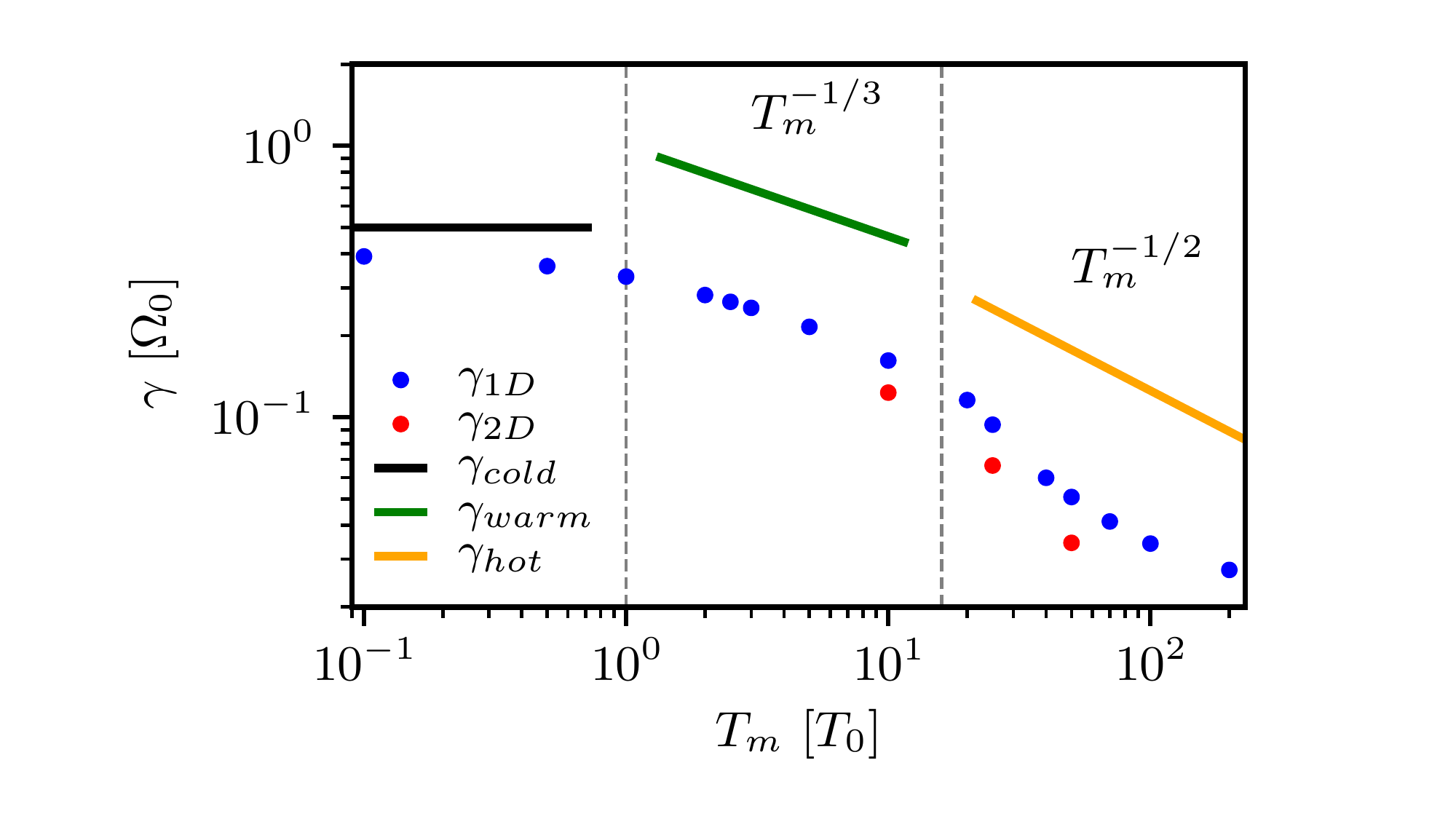}
    \caption{The linear growth rate of the non-resonant streaming instability under the effect of thermal effects. 1D (2D) run results are in blue (red). The solid line shows the linear theory growth rate in the cold (black), warm (green) and hot (yellow) cases. By courtesy of \cite{ref:mar}.}
    \label{F:NRHGENV}
\end{figure}

A recent analysis \cite{ref:mar} investigates the growth or the non-resonant branch and its saturation. The authors use an hybrid approach (electron treated as fluid, ions with a kinetic approach). They adopt the following set-up: Simulations are performed in 1D and 2D, the CR density is $n_{\rm CR}=0.01 n_b$, the CR drift is $u_{\rm CR} = 100 V_a$, 1000 macroparticles per cell (500 for thermal and 500 for non-thermal protons), the length unit is $\ell = c/\omega_{p,b}$ (the background proton inertial length), the simulation box is $L_x \times L_y = (1000, 200) \ell$. Boundary conditions are periodical. \\
Linear growth rate temperature dependences are well reproduced (see Fig. \ref{F:NRHGENV}). It appears that theoretical results exceed numerical ones by a factor of $\sim 2$. Theoretical results are obtained for the maximum growth rate while simulations sum all growing scales.\\

\begin{figure}
    \centering
    \includegraphics[width=1.\linewidth]{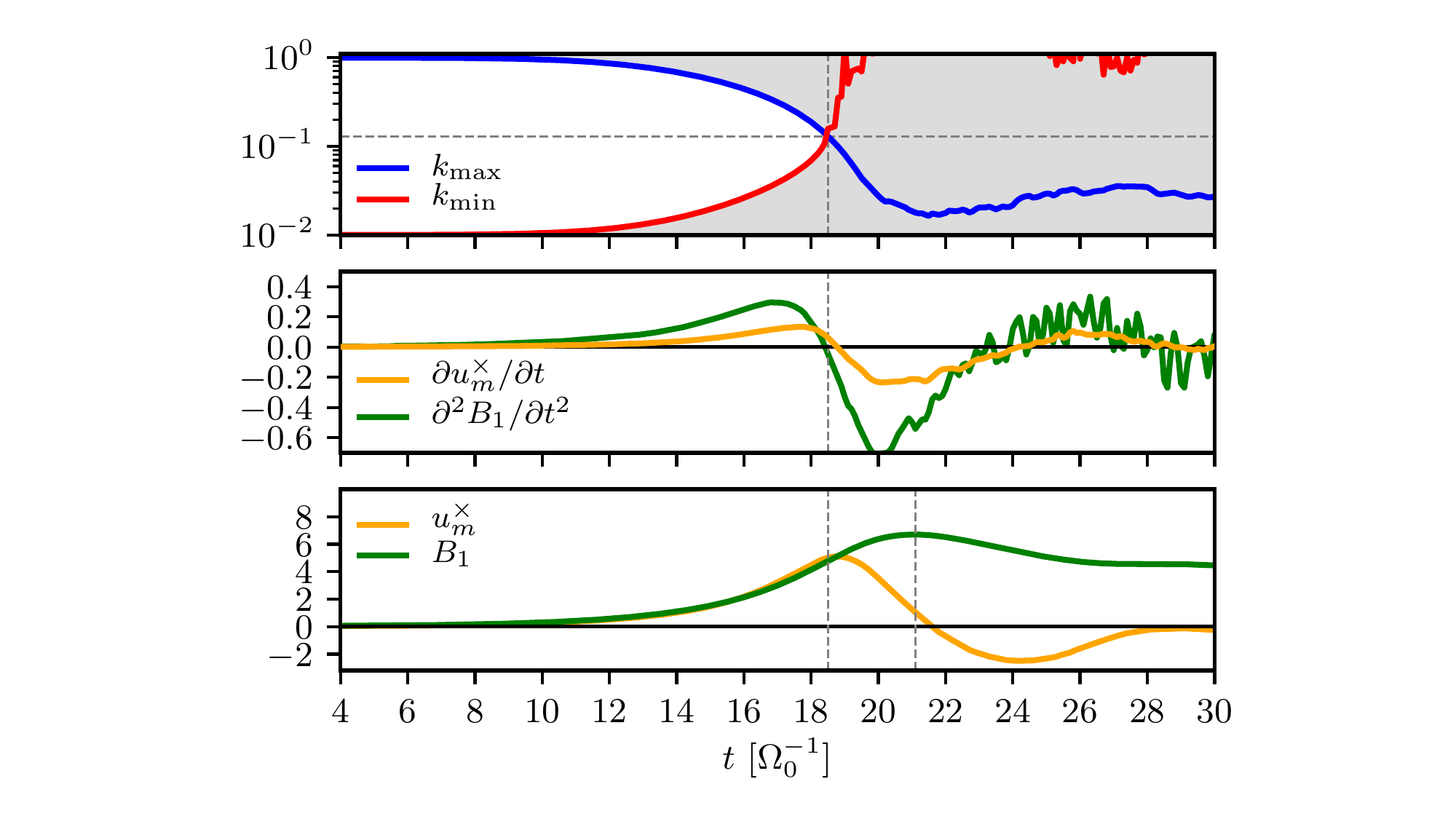}
    \caption{Upper panel: Time evolution of $k_1$ ($k_{\rm min}$ in the figure) and $k_2$ ($k_{\rm max}$in the figure) as function of time. maximum  The condition $k_1$ = $k_2$ is indicated with the vertical dashed line at t$\Omega \simeq$ 18.5 Greyed regions correspond to stable zones. Middle panel: background proton time derivative mean velocity parallel to the background B-field $u_m^x$ and perturbed magnetic field intensity second order
derivative.  Lower panel: perturbed magnetic field intensity $B_1$ and background protons normal fluid velocity. Magnetic saturation is indicated with
the vertical dashed line at $t \Omega \simeq$ 21. All Values are taken from
1D simulations. By courtesy of \cite{ref:mar}.}
    \label{F:SATNRH}
\end{figure}

Figure \ref{F:NRHGENV} shows the linear growth rate under the effect of the background plasma temperature $T_m$. 
The authors further analyse the saturation phase of the instability. Figure \ref{F:SATNRH} gives the time dependence of the magnetic energy and detail the origin of the magnetic saturation. It shows two distinct timescales. 
Rewritting the coupled Eqs between the perturbed gas velocity and magnetic field can help to understand the figure (see also Eqs. \ref{EQ:CRMHDl})

\begin{eqnarray}
 \partial_t \delta \vec{u} &=& {1 \over 4\pi \rho} \vec{B}_0.\vec{\nabla} \delta \vec{B} - {1 \over \rho} \vec{J}_{\rm CR} \wedge \delta \vec{B} \nonumber \\
 \partial_t \delta \vec{B} &=& \vec{B}_0.\vec{\nabla} \delta \vec{B} \ .\nonumber 
\end{eqnarray}

The first timescale at t$\Omega \simeq$ 18.5 marks the end of the linear phase and a transition to the non-linear growth phase. It is due to a change of sign of $\partial_t u_{1,\parallel}$, the time derivative of the parallel component of the gas velocity which is a balance between the magnetic tension (first term on RHS) and the Lorentz force (second term on RHS). Hence, this time marks the start of the dominance of the tension over the Lorentz force. In the meantime the second order magnetic perturbation derivative also changes of sign. At this time 
$k_1=k_2$. It seems that this condition is in fact the start of the non-linear growth phase not the saturation as primarily stated by \cite{ref:bell04}. The saturation appears a bit later at t$\Omega \simeq$ 21 where the sense of the parallel background velocity changes (see lower panel, yellow curve). In that case the electric field $\vec{E} = -{\vec{u} \over c} \wedge \vec{B}_0$ changes its sign and the CR are not anymore slowed down, which inhibit the magnetic field growth (see the green solid line in the lower panel).

\section{Conclusion}
Cosmic Rays are an important component of the interstellar medium even if they have very low densities as relativistic particles they carry a lot of momentum and pressure. This trend is even stronger in and/or nearby their sources where CR energy density can dominate gas thermal and magnetic energy densities. As so CRs are able to back react over the background ISM gas and on its macro-instabilities. In these lectures we discuss about one case: the Parker-Jeans instability, an instability thought to contribute to the galactic magnetic field dynamo. Then CRs can trigger their own instabilities because of their anisotropic distribution (streaming, pressured-instabilities) and/or because they carry a current as in the case of the non-resonant instability and its relatives instabilities. These lectures mostly focused on the derivation of the linear growth rate of these instabilities with a special focus on the two branches of the streaming instability. We also examined the effect of background plasma temperature and the case of partially ionised plasmas. These linear calculations serve as basis to numerical experiments which have the aim to evaluate the amplitude of magnetic fields produced via these instabilities in the shock and interstellar medium contexts, with the final aim of tracing the road to the turbulence generation, i.e. the redistribution of the energy from the forcing scale(s) towards many decades in fluctuations scale length, an essential step to a better constrain and understanding of CR propagation in our Galaxy. 
\acknowledgments
The author acknowledges discussions and comments by Anabella Araudo, Yoann Génolini, Ilya Plotnikov. The streaming instability growth rate derivation section is adapted from an original document by Xue Ning Bai.

\end{document}